\documentclass[journal,twoside,A4paper]{IEEEtran}

\usepackage{cite}
\usepackage[pdftex]{graphicx}
\DeclareGraphicsExtensions{.pdf,.eps}
\usepackage{amsmath}
\usepackage{amssymb}
\usepackage{caption}
\usepackage{subcaption}
\usepackage{hyperref}
\hypersetup{
pdfpagemode=UseNone,
pdfstartview=FitH,
pdftitle=Simple Thermal Noise Estimation of Switched Capacitor Circuits Based on OTAs -- Part II: SC Filters,
pdfauthor=Christian Enz,
pdfsubject=Noise in SC circuits,
pdfcreator=Christian Enz}
\usepackage[compress,capitalize]{cleveref}
\crefname{figure}{Fig.}{Figs.}
\crefname{equation}{}{}
\Crefname{equation}{Eq.}{Eqs.}
\crefname{secinapp}{Appendix}{appendices}
\Crefname{secinapp}{Appendix}{Appendices}

\newcounter{mytempeqncnt}

\newcommand{\super}[1]{\ensuremath{^{\mathrm{#1}}}}
\newcommand{\sub}[1]{\ensuremath{_{\mathrm{#1}}}}

\newcommand{\ELDO}{\textsc{{ELDO}\super{\copyright}}}

\newcommand{\TaH}{track \& hold} %

\newcommand{\Vout}{\ensuremath{V\sub{out}}}

\newcommand{\Gm}{\ensuremath{G_m}}
\newcommand{\Gon}{\ensuremath{G_{on}}}
\newcommand{\Ron}{\ensuremath{R_{on}}}

\newcommand{\kT}{\ensuremath{k_B T}}

\newcommand{\hfb}{\ensuremath{h_{fb}}}

\newcommand{\ph}[1]{\ensuremath{\Phi_#1}}
\newcommand{\phase}[1]{phase \ensuremath{\Phi_#1}}
\newcommand{\Phase}[1]{Phase \ensuremath{\Phi_#1}}

\newcommand{\Var}[2]{\ensuremath{#1_{n#2}^2}}
\newcommand{\Varp}[3]{\ensuremath{\left. #1_{n#2}^2 \right|_{\Phi_#3}}}
\newcommand{\Qnp}[2]{\ensuremath{\left. Q_{n#1} \right|_{\Phi_#2}}}
\newcommand{\Qn}[2]{\ensuremath{Q_{n #1}^2(#2)}}
\newcommand{\Varn}[3]{\ensuremath{#1_{n #2}^2(#3)}}

\newcommand{\Cinf}{\ensuremath{C_{\infty}}}
\newcommand{\Cinfn}[1]{\ensuremath{C_{\infty (#1)}}}
\newcommand{\Cinfp}{\ensuremath{C_{\infty}'}}
\newcommand{\Cinfpn}[1]{\ensuremath{C_{\infty (#1)}'}}
\newcommand{\Co}{\ensuremath{C_{0}}}
\newcommand{\Con}[1]{\ensuremath{C_{0 (#1)}}}
\newcommand{\aC}{\ensuremath{\alpha C}}
\newcommand{\Cin}{\ensuremath{C_{in}}}
\newcommand{\Cout}{\ensuremath{C_{out}}}
\newcommand{\CL}{\ensuremath{C_L}}
\newcommand{\Ceq}{\ensuremath{C_{eq}}}

\newcommand{\ain}{\ensuremath{\alpha_{in}}}
\newcommand{\aL}{\ensuremath{\alpha_L}}
\renewcommand{\a}{\ensuremath{\alpha}}

\newcommand{\Ts}{\ensuremath{T_s}}
\newcommand{\tset}{\ensuremath{t_{set}}}

\newcommand{\bota}{\ensuremath{\beta_{ota}}}
\newcommand{\bswi}{\ensuremath{\beta_{sw}}}
\newcommand{\botap}[1]{\ensuremath{\left.\beta_{ota} \right|_{\Phi_#1}}}
\newcommand{\bswip}[1]{\ensuremath{\left. \beta_{sw} \right|_{\Phi_#1}}}
\newcommand{\tota}{\ensuremath{\theta_{ota}}}
\newcommand{\tswi}{\ensuremath{\theta_{sw}}}
\newcommand{\tdir}{\ensuremath{\theta_{direct}}}
\newcommand{\totap}[1]{\ensuremath{\left.\theta_{ota} \right|_{\Phi_#1}}}
\newcommand{\tswip}[1]{\ensuremath{\left. \theta_{sw} \right|_{\Phi_#1}}}

\begin{document}

\title{Simple Thermal Noise Estimation of Switched Capacitor Circuits Based on OTAs -- Part II: SC Filters}

\author{\IEEEauthorblockN{
Christian Enz~\IEEEmembership{Fellow,~IEEE},
Sammy~Cerida~Rengifo~\IEEEmembership{Member,~IEEE},
Assim Boukhayma~\IEEEmembership{Member,~IEEE},\\
and
Fran\c{c}ois Krummenacher}

\thanks{Corresponding author: C. Enz (email: christian.enz@epfl.ch).}
\thanks{C. Enz and A. Boukhayma are with the Integrated Circuits Lab (ICLAB), Micro-engineering Institute, School of Engineering, EPFL.}
\thanks{Sammy~Cerida~Rengifo is with the CSEM SA, Neuch\^{a}tel}
\thanks{F. Krummenacher is with the Electrical Engineering Institute, School of Engineering, EPFL.}}

\markboth{Part II: SC Filters}%
{Enz \MakeLowercase{\textit{et al.}}:Simple Thermal Noise Estimation of SC Circuits Based on OTAs}

\maketitle
	
\begin{abstract}
In Part~I of this paper, we have shown how to calculate the thermal noise voltage variances in switched-capacitor (SC) circuits using operational transconductance amplifiers (OTAs) with capacitive feedback by using the extended Bode theorem. The method allows a precise estimation of the thermal noise voltage variances by simple circuit inspection without the calculation of any transfer functions nor integrals. While Part~I focuses on SC amplifiers and \TaH{} circuits, Part~II shows how to use the extended Bode theorem for SC filters. It validates the method on the basic integrator and then on a first-order low-pass filter by comparing the analytical results to transient noise simulations showing an excellent match.
\end{abstract}
	

\IEEEpeerreviewmaketitle
	
\section{Introduction}\label{sec:introduction}

An important class of switched-capacitor (SC) circuits are filters. They take advantage of the fact that the frequency characteristic only depends on capacitance ratios which turn out to be very accurate tanks to the excellent matching of capacitors and can be tuned by changing the clock frequency \cite{bib:gregorian:book:1986}. The calculation of noise in SC filters has always been difficult because SC circuits are linear time varying systems (LTVS) and therefore the noise calculation based on noise power spectral density (PSD) and transfer functions is tedious and impractical \cite{bib:schreier:tcas1:nov:2005}. Actually, most often the designer is not that much interested in the PSD, but rather the total noise power in the Nyquist band (from 0 to half the sampling frequency). In the early days, before advanced circuit simulators were available, specific tools were developed to calculate the PSD and the noise variance \cite{bib:goette:tcas:april:1989}, but they remained very complex and could not be used for simple hand calculation. Modern circuit simulators allow to compute noise for example in the time domain using transient noise analysis \cite{bib:bolcato:iscas:1992}. However, they don't provide simple analytical expressions of the noise voltage variance in order to optimize the SC filter noise.

Part~I demonstrated how the original Bode theorem, which is only valid for passive circuits, can be extended to active SC circuits made of capacitors, switches and operational transconductance amplifiers (OTAs) with capacitive feedback. It showed how the thermal noise voltage variance at any port of the SC circuit can be evaluated within each phase simply using three equivalent capacitances extracted by simple inspection from three different equivalent circuits. The methodology has then been applied to SC amplifiers and \TaH{} circuits and extensively validated with transient noise simulation. This Part~II of the paper shows how the extended Bode theorem can also be applied to SC filters realized with OTAs. The main difference between the examples of Part~I and SC filters is the fact that the virtual ground voltage controlling the OTA output current is now depending on several input voltages in addition to the OTA output voltage. Fortunately, in SC filters the contribution of these other voltage is much smaller than that of the OTA output voltage. Thanks to this property, it will be shown below that the extended Bode theorem can be applied.

The paper starts in \cref{sec:noise_mechanisms} with a qualitative description of the various noise mechanisms found in SC filters and shows how the extended Bode theorem can also be used for SC filters. \Cref{sec:verification on practical examples} presents its application to several practical examples starting with a passive first-order low-pass (LP) filter, then a stray-insensitive integrator, followed  by an active OTA-based first-order LP filter. The calculated noise in each case is compared with transient noise simulations showing an excellent match.

\section{Noise Mechanisms in SC Filters}\label{sec:noise_mechanisms}

\begin{figure}
  \centering
  \begin{subfigure}[t]{1\columnwidth}
    \centering
	\includegraphics[scale=0.8]{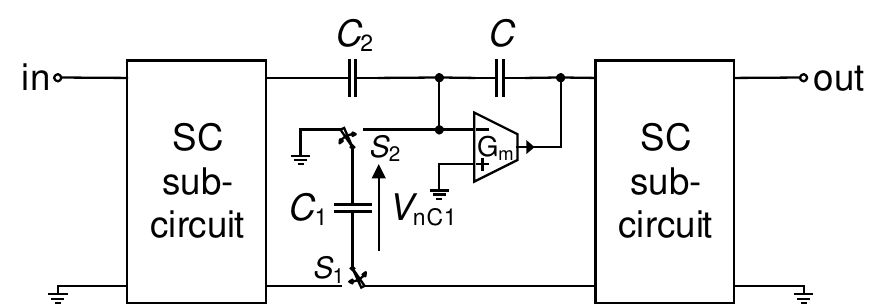}
	\caption{SC integrator during \phase{1}.}
	\label{fig:General_SC_circuit_phase1}
  \end{subfigure}%
  \\
  \begin{subfigure}[t]{1\columnwidth}
    \centering
    \includegraphics[scale=0.8]{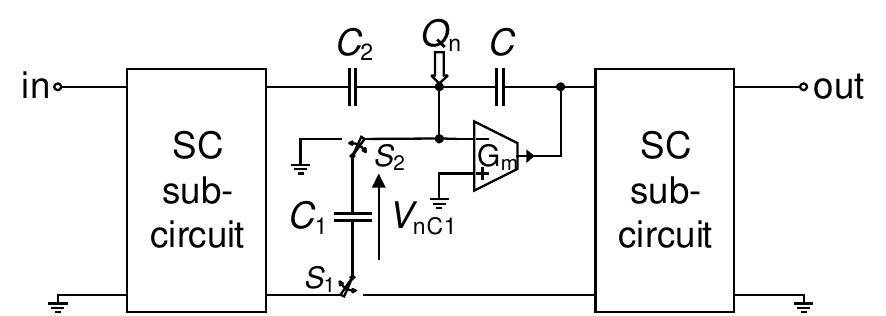}
    \caption{SC integrator during \phase{2}.}
    \label{fig:General_SC_circuit_phase2}
  \end{subfigure}
  \caption{SC integrator as part of a general SC circuit \cite{bib:enz:icnf:2015}.}
  \label{fig:General_SC_circuit}
\end{figure}

In this Section, we will recall the different noise mechanisms appearing in SC filters based on the stray-insensitive SC integrator shown in \Cref{fig:General_SC_circuit} as part of a larger SC filter. The integrator is made of an OTA having a transconductance \Gm, a feedback (or integrating) capacitor $C$, a switched-capacitor $C_1$ and an eventual capacitor $C_2$ always connected to the virtual ground. It is assumed that this SC filter uses two non-overlapping phases \ph{1} and \ph{2} which allows for the switches to be represented by toggle switches as shown in \cref{fig:General_SC_circuit}. It is also assumed that the OTA is ideal (in particular has infinite voltage gain and zero offset voltage) and that the circuit fully settles in each phase ($\Ron C \ll T/2$ and $\Ceq/\Gm \ll T/2$ where \Ceq{} is the equivalent capacitance accounting for the feedback). During each phase, the SC integrator operates as a continuous-time circuit. The noise is generated from the noise sources in the OTA and the switches. During the sampling \phase{1} (shown in \cref{fig:General_SC_circuit_phase1}), the noise voltage across the sampling capacitor $C_1$ is sampled at the end of \phase{1}, freezing a noise charge $\left.Q_n\right|_{\Phi_1}$ on $C_1$, which will be transferred to the integrating capacitor $C$ during the next \phase{2}. At the end of \phase{2} (shown in \cref{fig:General_SC_circuit_phase2}), not all of the charge stored on $C_1$ are transferred to the integrating capacitor $C$. Indeed, assuming that the OTA has infinite gain and has enough time to settle during \phase{2}, the voltage across $C_1$ at the end of \phase{2} should ideally be zero leaving no charge on $C_1$. However, because of the noise voltages coming from the switches and the OTA, the voltage across $C_1$ at the end of \phase{2} is not zero, leaving some random charge on $C_1$ that are not transferred to $C$. Because of charge conservation at the virtual ground node of the OTA, the random charge left on $C_1$ induces a random charge transfer error $\left.Q_n\right|_{\Phi_2}$ on capacitor $C$. The noise charge sampled on $C_1$ at the end of \phase{1} and transferred to the integrating capacitor $C$ during \phase{2} and the charge transfer error on the integrating capacitor due to the noise voltage across $C_1$ at the end of \phase{2}, can be modeled by a noise charge injector as shown in \cref{fig:General_SC_circuit_phase2}, injecting a noise charge $Q_n$ at the end of \phase{2}. The noise charges $\left.Q_n\right|_{\Phi_1}$ and $\left.Q_n\right|_{\Phi_2}$ are uncorrelated and the variance of the noise charge $Q_n$ injected into the virtual ground is therefore the sum of the variances of the noise charge due to the different noise sources active in each phase (switches and OTAs) and sampled at the end of phases $\Phi_1$ and $\Phi_2$
\begin{equation}\label{eqn:Q2n}
  \Var{Q}{} = \Varp{Q}{}{1} + \Varp{Q}{}{2}.
\end{equation}
The variances of the injected noise charge \Varp{Q}{}{1} and \Varp{Q}{}{2} are calculated in each phase from the variances of the voltages \Varp{V}{C_1}{1} and \Varp{V}{C_1}{2} as
\begin{equation}\label{eqn:General_SC_circuit_Q2C}
  \Varp{Q}{}{1} = C_1^2 \cdot \Varp{V}{C_1}{1}
  \quad \text{and} \quad
  \Varp{Q}{}{2} = C_1^2 \cdot \Varp{V}{C_1}{2}.
\end{equation}

This noise charge $Q_n$ is eventually shared with other capacitors connected in parallel to $C$ and then part of it is then propagating from the virtual ground where it was injected towards the filter output, resulting in a noise voltage at the output. The noise voltage variance at the filter output can be evaluated by first calculating the sampled-data $z$-transfer function from the noise charge injected at the virtual ground to the output using the technique described in \cite{bib:furrer:thesis:1983,bib:enz:icnf:2015}. This technique is general but quite cumbersome and we will show below that in some simple cases, it can also be calculated directly in the time domain using recursive relations.

\begin{figure}
  \centering
  \begin{subfigure}[t]{1\columnwidth}
    \centering
	\includegraphics[scale=0.8]{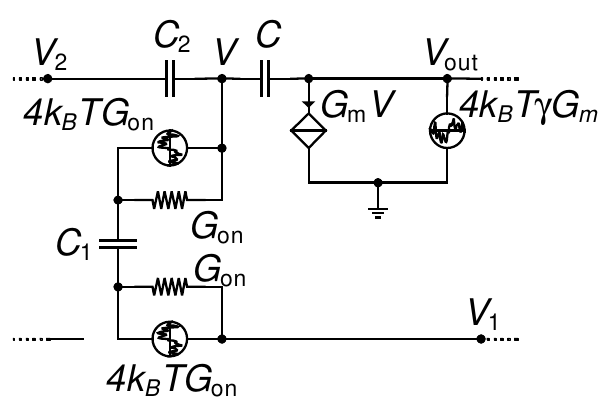}
	\caption{Equivalent linear schematic of the integrator of \cref{fig:General_SC_circuit}.}
	\label{fig:General_SC_equivalent_circuit1}
  \end{subfigure}%
  \\
  \begin{subfigure}[t]{1\columnwidth}
    \centering
    \includegraphics[scale=0.8]{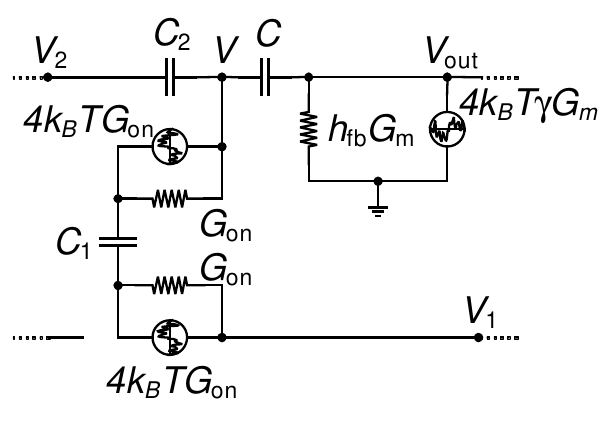}
    \caption{Simplified equivalent linear schematic.}
    \label{fig:General_SC_equivalent_circuit2}
  \end{subfigure}
  \caption{Equivalent schematics of the SC circuit of \cref{fig:General_SC_circuit} \cite{bib:enz:icnf:2015}.}
  \label{fig:General_SC_equivalent_circuit}
\end{figure}

According to \cref{eqn:General_SC_circuit_Q2C}, in order to calculate the noise charge variances, we need to compute the noise voltage variances across all the switched capacitors connected to the virtual ground of the integrator. This can be done by calculating the PSD of the noise voltage across the switched capacitors during each phase and integrating it over frequency. Obtaining these PSD requires to first derive the transfer functions from the noise sources (switches and OTAs) to the voltage across the switched capacitors. The noise voltage variances are finally obtained by integrating the PSD over frequency. All this process is actually a quite fastidious task. Alternatively, we can check whether the extended Bode theorem presented in Part~I of this article may eventually be used to this purpose. Contrary to the SC amplifier and \TaH{} circuits discussed in Part~I, in the case of SC filters, the voltage at the virtual ground of the OTA shown in \cref{fig:General_SC_circuit} does not only depend on the output voltage \Vout{} but also on the voltages $V_1$ and $V_2$.

In order to calculate the effective virtual ground voltage $V$, the integrator shown in \cref{fig:General_SC_circuit}, can be modelled by the equivalent linear circuit shown in \cref{fig:General_SC_equivalent_circuit1}. The switches are replaced by their on-conductance $\Gon$ in parallel with their thermal noise current source having a PSD $4\kT \Gon$. The OTA is modelled by a voltage-controlled current source (VCCS) in parallel with its thermal noise current source having a PSD given by $4\kT \gamma \Gm$ where $\gamma$ is the OTA noise excess factor defined as $\gamma \triangleq G_m \cdot R_{nth}$ where $R_{nth}$ is the OTA input-referred thermal noise resistance. From the circuit of \cref{fig:General_SC_equivalent_circuit1}, the voltage $V$ at the virtual ground node can be expressed as
\begin{equation}\label{eqn:V}
  V =  \beta_1 (\omega) \cdot V_1 +  \beta_2 (\omega) \cdot V_2 + \hfb(\omega) \cdot V_{out}.
\end{equation}

\begin{figure*}[!hbt]
  \centering
  \begin{subfigure}[t]{.33\textwidth}
    \centering
	\includegraphics[scale=.8]{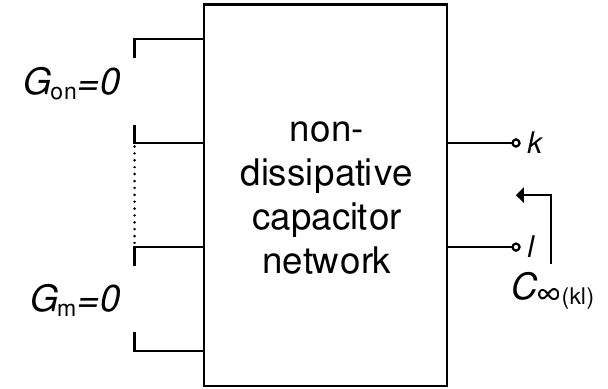}
	\caption{Equivalent circuit used for the calculation of \Cinfn{kl}: all switches and OTAs of the SC circuit are removed.}
	\label{fig:Equivalent_circuit_Cinf}
  \end{subfigure}%
  \begin{subfigure}[t]{.33\textwidth}
    \centering
	\includegraphics[scale=.8]{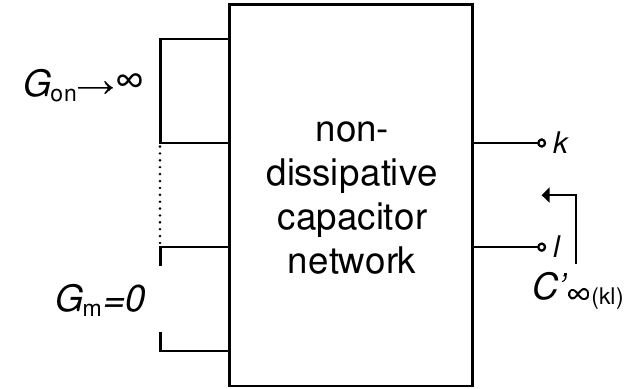}
	\caption{Equivalent circuit used for the calculation of \Cinfpn{kl}: all switches that are closed during the clock phase in consideration are replaced by short-circuits and all OTAs of the SC circuit	are removed.}
	\label{fig:Equivalent_circuit_Cinf_prime}
  \end{subfigure}
  \begin{subfigure}[t]{.33\textwidth}
    \centering
	\includegraphics[scale=.8]{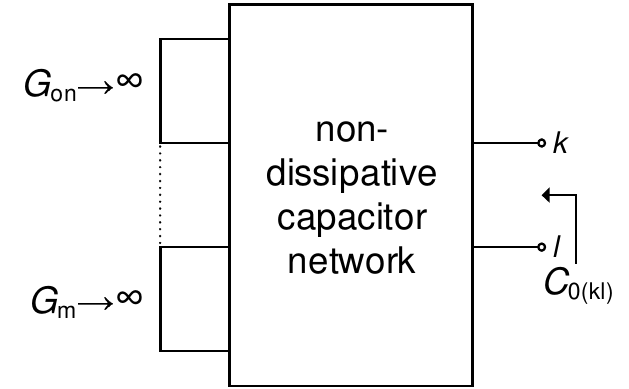}
	\caption{Equivalent circuit used for the calculation of \Con{kl}: all switches that are closed during the clock phase in consideration are replaced by short-circuits and all OTAs of the SC circuit have their output shorted to ground.}
	\label{fig:Equivalent_circuit_C0}
  \end{subfigure}
  \caption{Capacitances calculation for the extended Bode theorem \cite{bib:enz:icnf:2015}.}
  \label{fig:Equivalent_circuit_C}
\end{figure*}

Assuming that the switch conductance \Gon{} is very large (more precisely that $\Gon \gg 4C_1(C+C_2)/T$), the transfer functions $\beta_1(\omega)$, $\beta_2(\omega)$ and the feedback voltage gain $\hfb(\omega)$ can be assumed frequency-independent and are given by
\begin{subequations}
  \begin{align}
    \beta_1 &= \frac{C_1}{C + C_1 + C_2} \cong \frac{C_1}{C},\label{eqn:beta1}\\
    \beta_2 &= \frac{C_2}{C + C_1 + C_2} \cong \frac{C_2}{C},\label{eqn:beta2}\\
    \hfb &= \frac{C}{C + C_1 + C_2} \cong 1,\label{eqn:hfb}
  \end{align}
\end{subequations}
where the approximations account for the fact that in SC filters, contrary to the cases of the SC amplifier and \TaH{} circuits analyzed in Part~I, the integrator feedback capacitance $C$ is usually much larger than the switched and non-switched capacitances $C_1$ and $C_2$ ($C_1, C_2 \ll C$). Additionally, capacitors $C_1$ and $C_2$ connected to the virtual ground, have their other node connected either to the ground or to the output of another OTA. Hence, voltages $V_1$, $V_2$ and $V_{out}$ in \cref{eqn:V} are of the same order of magnitude and ultimately bounded by the supply voltage $V_{DD}$. Under the above assumptions, the two first terms in \cref{eqn:V} can be neglected and \cref{eqn:V} can be approximated by
\begin{equation}\label{eqn:V_approx}
  V \cong \hfb \cdot V_{out}.
\end{equation}

Using \cref{eqn:V_approx}, the VCCS can now be replaced by a conductance of value $\hfb \cdot \Gm$ leading to the equivalent circuit shown in \cref{fig:General_SC_equivalent_circuit2}. As in Part~I, the latter circuit can be considered as a passive $RC$ network and the extended Bode theorem can be used. The equation to calculate the thermal noise voltage variance between any node $k$ and $l$ of the SC filter is recalled here because it is central to the calculation method
\begin{equation}\label{eqn:V2n_Bode_OTA}
  \Var{V}{(kl)} = \kT \cdot \left[\frac{1}{\Cinfn{kl}} + \frac{\gamma/\hfb - 1}{\Cinfpn{kl}}
  - \frac{\gamma/\hfb}{\Con{kl}}\right].
\end{equation}

\Cref{eqn:V2n_Bode_OTA} only requires the evaluation of the three capacitances \Cinfn{kl}, \Cinfpn{kl} and \Con{kl}. The latter can easily be calculated by inspection of the three equivalent circuits depicted in \cref{fig:Equivalent_circuit_C} which are each composed only of capacitors. The extended Bode theorem will now be illustrated and validated by transient noise simulations for various SC filters in the next Section.

\section{Practical Examples of Thermal Noise Estimation in SC Filters}\label{sec:verification on practical examples}

SC filters operate periodically and each period corresponds to a succession of phases and, in each phase, the circuit corresponds to a continuous-time passive or OTA-based capacitive circuit. Hence, the variance of the noise charge held in each capacitor of the SC filter at the end of each phase can be estimated using either the original Bode theorem for passive circuits or the extended Bode theorem presented in the previous Section for OTA-based active SC filters. These thermal noise charges generated in different phases are uncorrelated in time. Thus, the total noise charge held on a capacitor $C$ after one switching period can be expressed as the sum of the noise charges generated in each phase. Unlike the SC amplifier and \TaH{} examples presented in Part~I, where the capacitors are reset between two successive periods, in SC filters the capacitors (particularly the integrating capacitor) are not always reset. A recursive relation can then be established between the noise variance at the end of the present switching period and the previous ones.

For all the SC filters analyzed in this Section, it is assumed that the circuit operates with two non-overlapping phases \ph{1} and \ph{2} which allows for the switches to be represented by toggle switches. It is also assumed that the OTAs have infinite voltage gain, have no offset, are linear, don't show any slew-rate and can therefore be represented by simple VCCS. It is also assumed that the time constants related to the switches are much smaller than half a period ($R_{on} C \ll T/2$) and that the OTA output node fully settles within each phase. This requires the settling time $\tset=\Ceq/\Gm$ to be much smaller than half the sampling period $\Ts/2$ where $\Ceq=\Cout/\beta$ and $\Cout=\CL+(1-\beta)C_2$ is the effective load capacitance with $\beta=C_2/(C_1+C_2+\Cin)$ the feedback gain.

The validation of the thermal noise estimation method is performed by transient noise simulation using the \ELDO{} simulator \cite{bib:bolcato:iscas:1992}. The transient noise simulation technique is close to the physical noise behavior taking place in SC circuits and is quite easy to set up. It is also the most suitable for verifying the noise behavior through the phases and switching periods for the different simulated filters. The transient noise simulations are performed using ideal components, in particular the OTAs are modelled by simple VCCS with a thermal noise current source at the output having a PSD $4 \kT \gamma \Gm$. In order to obtain reliable results, the simulator noise bandwidth is set to be higher than the maximum pole of all the transfer functions seen by any noise source in the circuit during any phase.

The validation starts with the passive first-order LP filter which noise voltage variances in each phase can actually be calculated by means of the original Bode theorem since it is a passive circuit. The second example is the basic stray-insensitive SC integrator which is the main building block for implementing higher order SC filters. Finally, the first-order active LP filter is analyzed.

\subsection{Passive First-order LP Filter}

\begin{figure}[!h]
  \centering
  \begin{subfigure}[t]{0.49\columnwidth}
    \centering
    \includegraphics[scale=0.8]{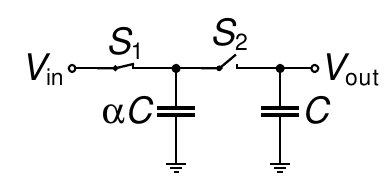}
    \caption{\centering \Phase{1}.}
    \label{fig:SC_LP_Pas_ph1}
  \end{subfigure}
  \begin{subfigure}[t]{0.49\columnwidth}
    \centering
    \includegraphics[scale=0.8]{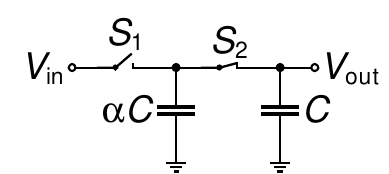}
    \caption{\centering \Phase{2}.}
    \label{fig:SC_LP_Pas_ph2}
  \end{subfigure}
  \caption{Passive SC first-order LP filter.}
  \label{fig:SC_LP_Pas}
\end{figure}

\begin{figure*}[!t]
  \centering
  \begin{subfigure}[t]{.32\textwidth}
    \centering
    \includegraphics[scale=0.8]{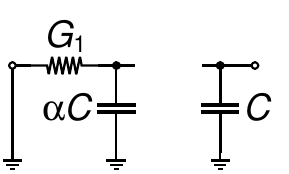}
    \caption{\centering Equivalent passive SCLPF circuit during \phase{1}.}
    \label{fig:SC_LP_Pas_ph1a}
  \end{subfigure}
  \begin{subfigure}[t]{.32\textwidth}
    \centering
    \includegraphics[scale=0.8]{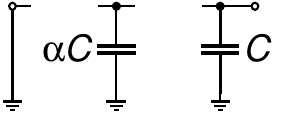}
    \caption{\centering $\Cinfn{\aC}=\aC$ }
    \label{fig:SC_LP_Pas_ph1b}
  \end{subfigure}
  \begin{subfigure}[t]{.32\textwidth}
    \centering
    \includegraphics[scale=0.8]{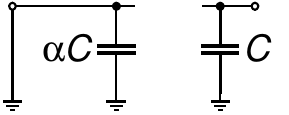}
    \caption{\centering $\Con{\aC}=\infty$}
    \label{fig:SC_LP_Pas_ph1c}
  \end{subfigure}
  \caption{Equivalent circuit schematics for noise variance calculation in \phase{1}.}
  \label{fig:SC_LP_Pas_ph1abc}
\end{figure*}

\begin{figure*}[!t]
  \centering
  \begin{subfigure}[t]{.32\textwidth}
    \centering
    \includegraphics[scale=0.8]{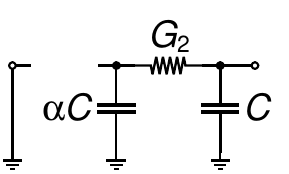}
    \caption{\centering Equivalent passive SCLPF circuit during \phase{2}.}
    \label{fig:SC_LP_Pas_ph2a}
  \end{subfigure}
  \begin{subfigure}[t]{.32\textwidth}
    \centering
    \includegraphics[scale=0.8]{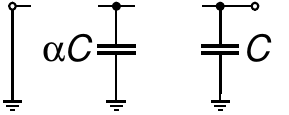}
    \caption{\centering $\Cinfn{C}=C$}
    \label{fig:SC_LP_Pas_ph2b}
  \end{subfigure}
  \begin{subfigure}[t]{.32\textwidth}
    \centering
    \includegraphics[scale=0.8]{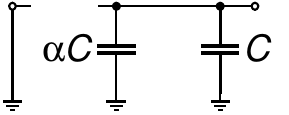}
    \caption{\centering $\Con{C}=\aC+C$}
    \label{fig:SC_LP_Pas_ph2c}
  \end{subfigure}
  \caption{Equivalent circuit schematics for noise variance calculation in \phase{2}.}
  \label{fig:SC_LP_Pas_ph2abc}
\end{figure*}

\cref{fig:SC_LP_Pas} shows the schematic of a passive first-order LP filter where the SC $\aC$ plays the role of the resistance of a simple first-order LP $RC$ filter. The circuit is operated periodically in two phases. During \phase{1}, the input voltage is sampled on capacitor $\aC$, while the voltage across $C$ is read out. During \phase{2}, the charge sampled on $\aC$ is then added to the charge already existing on $C$ and the total charge on $\aC$ and $C$ is then shared between $\aC$ and $C$. The accumulation and averaging of charge occurring during \phase{2} translates into a recursive relation between the output and input voltages resulting in a transfer function in the $z$-domain given by
\begin{equation}\label{eqn:SC_LP_Pas_transfer_function}
  H(z)=\frac{\alpha z^{-1}}{1+\alpha-z^{-1}}.
\end{equation}

For frequencies much smaller than the sampling frequency, this circuit operates as a first-order LP filter with a cutoff frequency given by
\begin{equation}\label{eqn:SC_LP_Pas_cutoff}
  f_c=\frac{\alpha}{1+\alpha} \cdot \frac{f_s}{2\pi},
\end{equation}
where $f_s$ is the sampling frequency.

The calculation of the output noise voltage variance is detailed below. During \phase{1}, the noise generated by the input switch S\sub{1} is first sampled on capacitor $\aC$. The noise voltage variance across capacitor $\aC$ during \phase{1} can be calculated by applying the Bode theorem as explained in Part~I with the capacitances \Cinf{} and \Co{} shown in \cref{fig:SC_LP_Pas_ph1abc} and resulting in
\begin{equation}\label{eqn:SC_LP_Pas_V_C_ph1}
  \Varp{V}{\aC}{1} = \kT \cdot \left[\frac{1}{\aC} - 0 \right] = \frac{\kT}{\aC}.
\end{equation}
The noise charge frozen on capacitor $\aC$ at the end of \phase{1} can hence be expressed as
\begin{equation}\label{eqn:SC_LP_Pas_Q_ph1}
  \Varp{Q}{\aC}{1} = \kT \cdot \aC.
\end{equation}
This charge $\left. Q_{n\aC}\right|_{\Phi_1}$ is then shared between capacitors $\aC$ and $C$ during \phase{2}. At the end of \phase{2}, after the switch S\sub{2} has opened, only a fraction $1/(1+\alpha)$ of the charge $\left. Q_{n\aC} \right|_{\Phi_1}$ will remain on capacitor $C$. Hence the variance of the noise charge generated during \phase{1} and remaining on capacitor $C$ at the end of \phase{2} is given by
\begin{equation}\label{eqn:SC_LP_Pas_Q_C_ph1}
  \Varp{Q}{C}{1} = \kT \cdot \aC \cdot \left(\frac{1}{1+\alpha}\right)^2 = \frac{\kT \cdot \aC}{(1+\alpha)^2}.
\end{equation}

To this charge generated during \phase{1} and left on $C$ during \phase{2} adds the charge sampled on $C$ at the end of \phase{2} due to the noise generated by the switch S\sub{2}. This noise charge can be calculated directly using the Bode theorem with the capacitances \Cinf{} and \Co{} shown in \cref{fig:SC_LP_Pas_ph2abc} for \phase{2} resulting in
\begin{equation}\label{eqn:SC_LP_Pas_V_CL_ph2}
  \Varp{V}{C}{2} = \kT \cdot \left[\frac{1}{C} - \frac{1}{C+\aC}\right]
  = \frac{\kT}{C} \cdot \frac{\alpha}{1+\alpha}.
\end{equation}
The corresponding noise charge sampled at the end of \phase{2} on capacitor $C$ and due to the noise generated by switch S\sub{2} simplifies to
\begin{equation}\label{eqn:SC_LP_Pas_Q_ph2}
  \Varp{Q}{C}{2} = \kT \cdot C \cdot \frac{\alpha}{1+\alpha}.
\end{equation}
The variance of the total noise charge injected on $C$ at the end of \phase{2} due to the noise generated during \phase{1} and \phase{2} is then given by
\begin{equation}\label{eqn:SC_Pas_Q2n}
  \Var{Q}{} = \Varp{Q}{C}{1} + \Varp{Q}{C}{2} = \kT \cdot \aC \cdot \frac{2+\alpha}{(1+\alpha)^2}.
\end{equation}

Capacitor $C$ is not reset between two consecutive periods. Hence, the noise charge $Q_n$ generated during the two phases of a current period, which variance is calculated above, adds to the noise charge already held on capacitor $C$ and originated during the previous switching periods. Let $Q_{nC}(n)$ be the noise charge held on capacitor $C$ at the end of the $n^{th}$ switching period. The latter charge is held on $C$ during \phase{1} of the next switching period $n+1$ and then shared between $\aC$ and $C$ at the beginning of \phase{2}. When switch S\sub{2} opens at the end of \phase{2}, a fraction $1/(1+\alpha)$ of this charge will remain on $C$. The variance of the total noise charge sampled in $C$ at the end of the $(n+1)^{th}$ switching period can then be expressed as
\begin{equation}\label{eqn:SC_LP_Pas_ch_eq}
  \Varn{Q}{C}{n+1} = \frac{\Varn{Q}{C}{n}}{(1+\alpha)^2} + \Var{Q}.
\end{equation}

Since \cref{eqn:SC_LP_Pas_ch_eq} is a recursive relation, the noise charge variance at the end of the $n^{\text{th}}$ period can be expressed as
\begin{equation}\label{eqn:SC_LP_Pas_V_CL_conv}
  \Varn{Q}{C}{n} = \kT \cdot C \cdot \left[1- \left(\frac{1}{1+\alpha}\right)^{2n}\right].
\end{equation}

After several switching periods, depending on the capacitors ratio, the second term in \cref{eqn:SC_LP_Pas_V_CL_conv} tends to unity and the noise voltage variance seen across $C$ converges to
\begin{equation}\label{eqn:SC_LP_Pas_sampled}
  \Var{V}{C} = \frac{\kT}{C}.
\end{equation}
This result might seem trivial since it actually corresponds to the variance of the noise voltage of the equivalent continuous-time first-order LP $RC$ filter! However, the continuous-time equivalent circuit of SC circuits does not always provide the correct thermal noise variance.

\begin{figure}[!t]
  \centering
  \begin{subfigure}[t]{1\columnwidth}
    \centering
    \includegraphics[width=0.8\columnwidth]{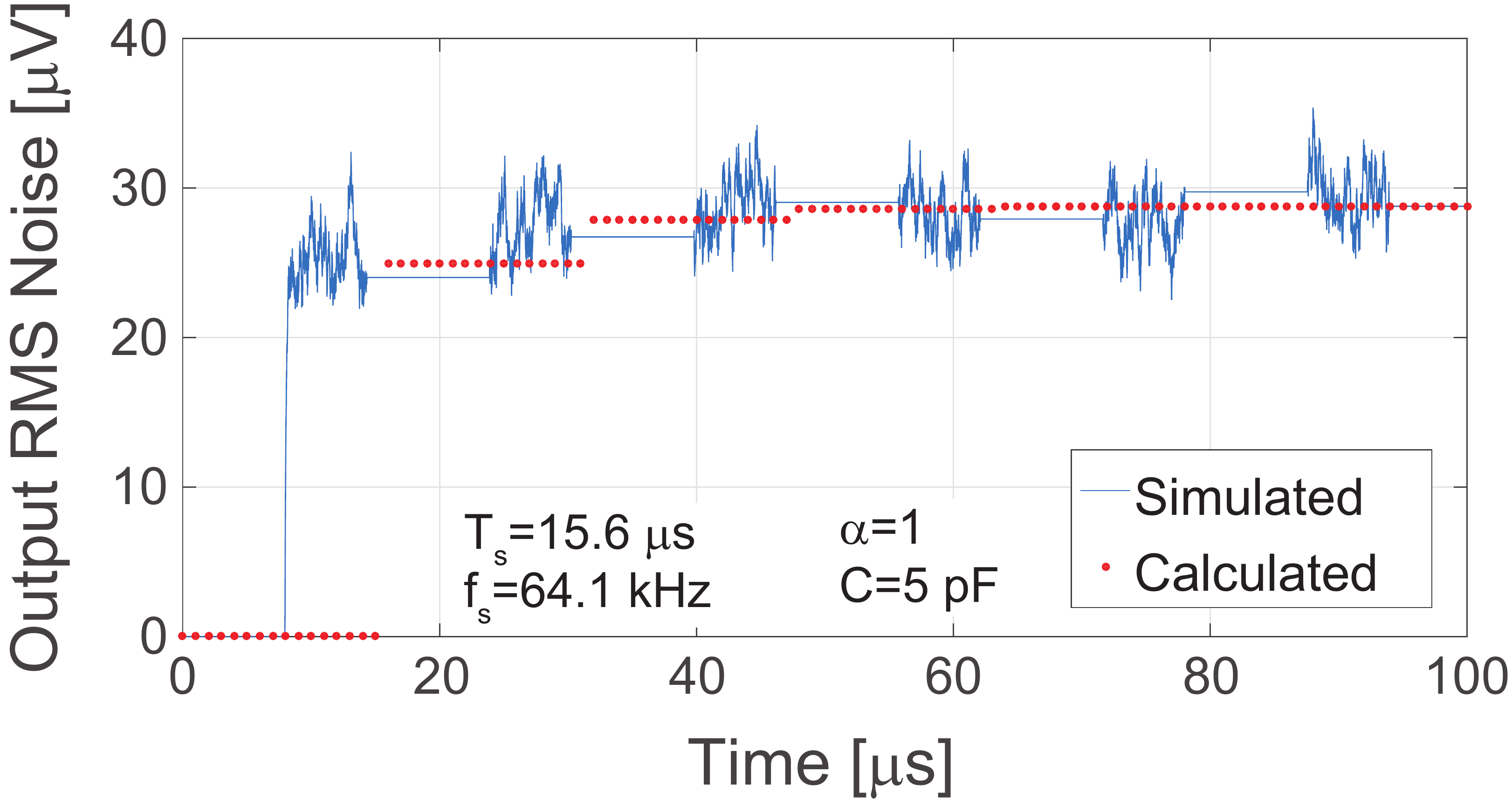}
	\caption{\centering $C=\aC=5\;pF$ ($\alpha=1$) leading to $V_{nC}=28.8\;\mu V_{rms}$ at $T=300\,K$.}
	\label{fig:SC_LP_Pas_trans_a}
  \end{subfigure}%
  \\
  \begin{subfigure}[t]{1\columnwidth}
    \centering
	\includegraphics[width=0.8\columnwidth]{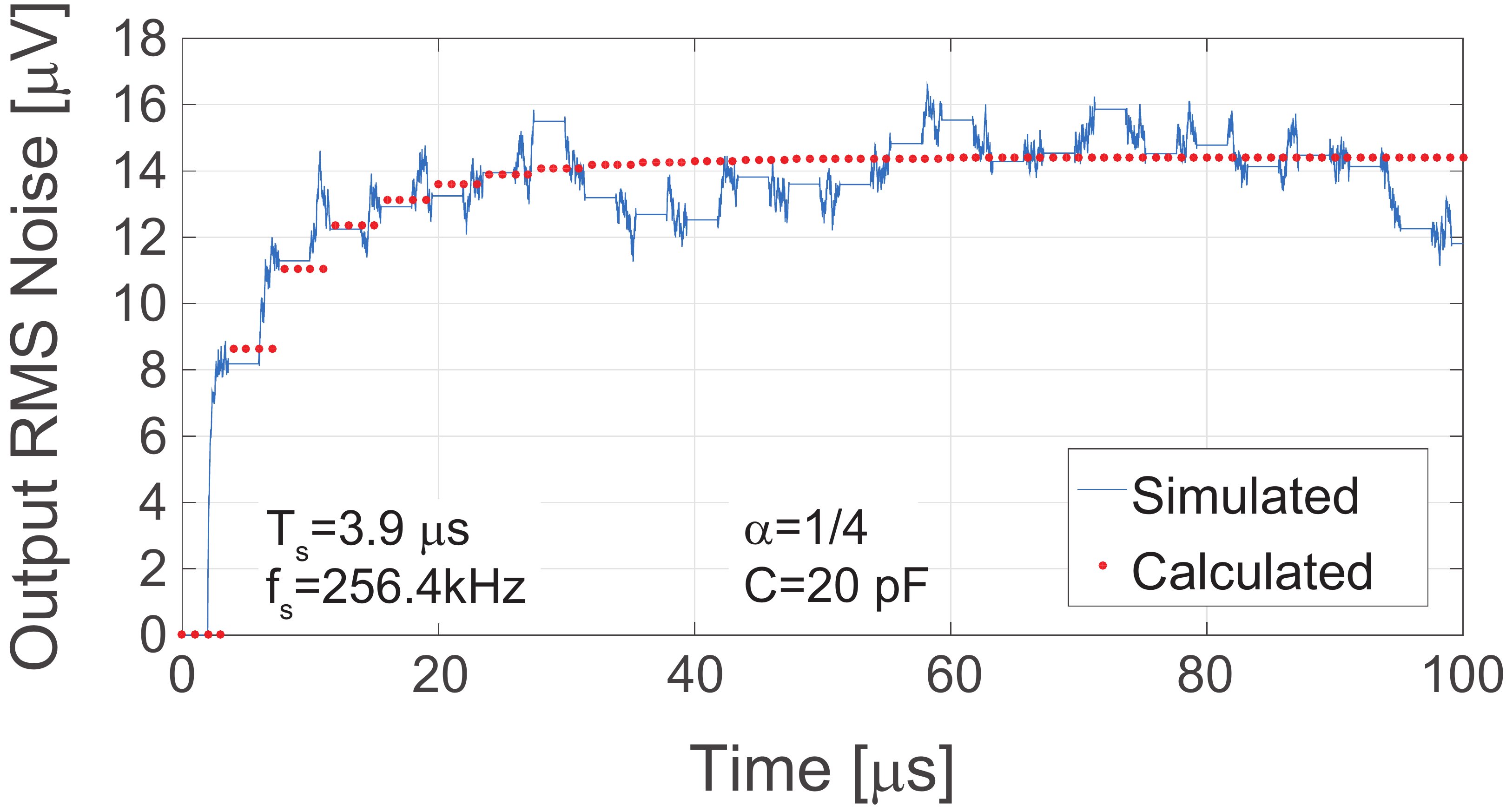}
	\caption{\centering $C=4 \aC=20\;pF$ ($\alpha=1/4$) leading to $V_{nC}=14.4\;\mu V_{rms}$ at $T=300\,K$.}
	\label{fig:SC_LP_Pas_trans_b}
  \end{subfigure}
  \caption{Simulated transient noise output RMS voltage compared to calculated for a passive SCLPF for two different values of capacitances.}
  \label{fig:SC_LP_trans}
\end{figure}

In order to validate the above noise estimation, the circuit is simulated with capacitors $\aC=C=5\;pF$, as shown in \cref{fig:SC_LP_Pas_trans_a}. The output noise is calculated for the readout phase (\ph{1}) when the output is floating and before the injection (\ph{2}), hence the calculated value remains constant for the entire period. The value of the capacitors chosen leads to a rapid convergence, it only takes $3$ periods. The transient noise simulation validates the noise estimation method for each period of sampling and confirms than the noise increases through the periods and converges to a constant value as predicted by the calculation based on the presented noise estimation method.

\cref{fig:SC_LP_Pas_trans_b} shows a simulation with a different capacitance ratio $\alpha$ chosen for a slower convergence in order to validate the noise convergence in detail. The recursive relation and the convergence of the output noise is confirmed by the perfect match of the simulated and calculated results. These calculations also agree with the results presented in \cite{bib:boukhayma:newcas:2015}.

\subsection{The Integrator}

\begin{figure}[!ht]
  \centering
  \begin{subfigure}[t]{0.49\columnwidth}
    \centering
    \includegraphics[scale=0.8]{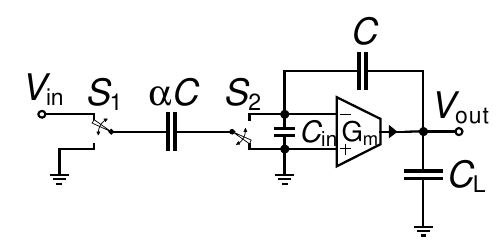}
    \caption{\centering \Phase{1}.}
    \label{fig:SC_int_ph1}
  \end{subfigure}
  \begin{subfigure}[t]{0.49\columnwidth}
    \centering
    \includegraphics[scale=0.8]{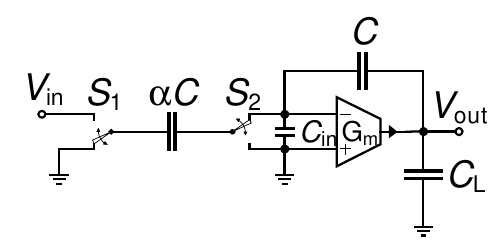}
    \caption{\centering \Phase{2}.}
    \label{fig:SC_int_ph2}
  \end{subfigure}
  \caption{Stray-insensitive non-inverting SC integrator \cite{bib:gregorian:book:1986}.}
  \label{fig:SC_int}
\end{figure}

\cref{fig:SC_int} shows the implementation of a non-inverting stray-insensitive SC integrator \cite{bib:gregorian:book:1986} where $\alpha$ is the ratio between the sampling capacitance $\aC$ and the integrating capacitance $C$. Each switching period of the SC integrator is composed of two non-overlapping phases. A charge is sampled on capacitor $\aC$ at the end of \phase{1} while the integrator output voltage is read by the next stage. This charge is then transferred to $C$ during \phase{2} and adds to the charge already held on $C$ from the previous period. The transfer function of this stage in the $z$-domain is given by \cite{bib:gregorian:book:1986}
\begin{equation}\label{eqn:SC_int_transfer_function}
  H(z)=\frac{\alpha}{1-z^{-1}}.
\end{equation}

\begin{figure*}[!hbt]
  \centering
  \begin{subfigure}[t]{.24\textwidth}
    \centering
    \includegraphics[scale=.6]{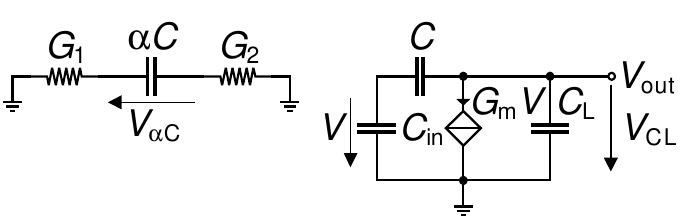}
    \caption{\centering Equivalent SC integrator circuit during \phase{1}.}
    \label{fig:SC_int_ph1a}
  \end{subfigure}
  \begin{subfigure}[t]{.24\textwidth}
    \centering
    \includegraphics[scale=.6]{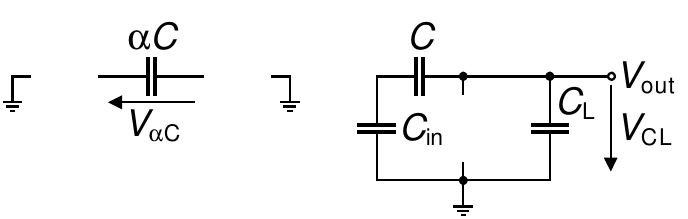}
    \caption{\centering $\Cinfn{\aC}=\aC$\\
    $\Cinfn{\CL}=\CL+\frac{\Cin C}{\Cin+C}$.}
    \label{fig:SC_int_ph1b}
  \end{subfigure}
  \begin{subfigure}[t]{.24\textwidth}
    \centering
    \includegraphics[scale=.6]{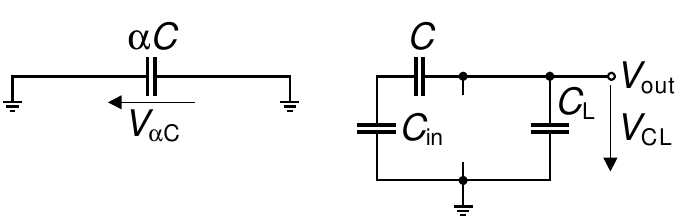}
    \caption{\centering $\Cinfpn{\aC}=\infty$\\
    $\Cinfpn{\CL}=\CL+\frac{\Cin C}{\Cin+C}$.}
    \label{fig:SC_int_ph1c}
  \end{subfigure}
  \begin{subfigure}[t]{.24\textwidth}
    \centering
    \includegraphics[scale=.6]{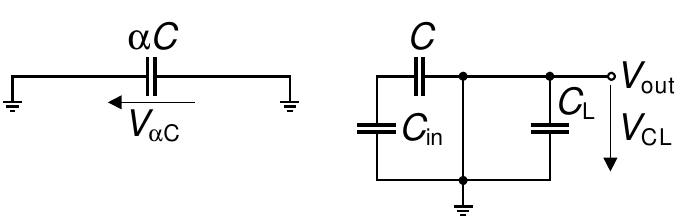}
    \caption{\centering $\Con{\aC}=\infty$\\
    $\Con{\CL}=\infty$.}
    \label{fig:SC_int_ph1d}
  \end{subfigure}
  \caption{Equivalent circuit schematics for noise variance calculation in \phase{1}.}
  \label{fig:SC_int_ph1abcd}
\end{figure*}

\begin{figure*}[!hbt]
  \centering
  \begin{subfigure}[t]{.24\textwidth}
    \centering
    \includegraphics[scale=.6]{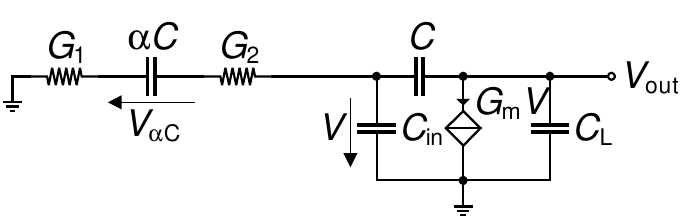}
    \caption{\centering Equivalent SC integrator circuit during \phase{2}.}
    \label{fig:SC_int_ph2a}
  \end{subfigure}
  \begin{subfigure}[t]{.24\textwidth}
    \centering
    \includegraphics[scale=.6]{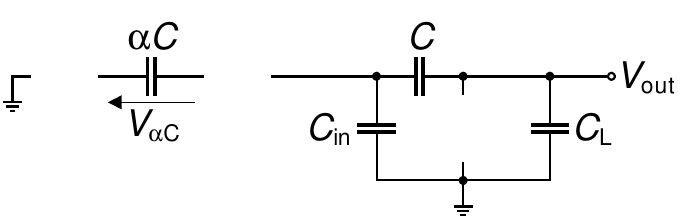}
    \caption{\centering $\Cinfn{\aC}=\aC$.}
    \label{fig:SC_int_ph2b}
  \end{subfigure}
  \begin{subfigure}[t]{.24\textwidth}
    \centering
    \includegraphics[scale=.6]{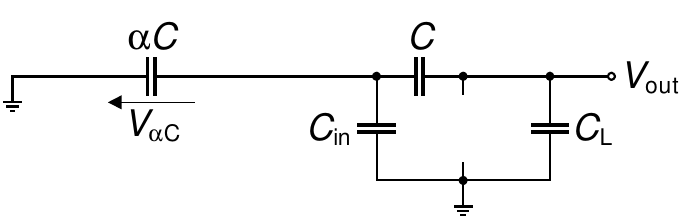}
    \caption{\centering $\Cinfpn{\aC}=\aC+\Cin+\frac{C \CL}{C+\CL}$.}
    \label{fig:SC_int_ph2c}
  \end{subfigure}
  \begin{subfigure}[t]{.24\textwidth}
    \centering
    \includegraphics[scale=.6]{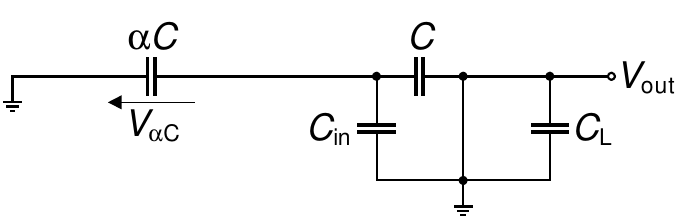}
    \caption{\centering $\Con{\aC}=\aC+\Cin+C$.}
    \label{fig:SC_int_ph2d}
  \end{subfigure}
  \caption{Equivalent circuit schematics for noise variance calculation in \phase{2}.}
  \label{fig:SC_int_ph2abcd}
\end{figure*}

The goal is to calculate the variance of the thermal noise voltage at the output. During \phase{1}, the only capacitor sampling a noise charge is $\aC$, since the other capacitances are holding their charge sampled in the previous period (except for the part that comes from the direct noise which will be accounted for when calculating the noise during \phase{2}). The noise voltage variance across capacitor $\aC$ during \phase{1} is calculated using the extended Bode theorem \cref{eqn:V2n_Bode_OTA} with the help of the schematics shown in \cref{fig:SC_int_ph1abcd} for the calculation of capacitances \Cinf{}, \Cinfp{} and \Co{}. Since the OTA is completely disconnected from the sampling capacitor $\aC$ during \phase{1}, it does not contribute to the noise sampled on capacitor $\aC$ at the end of \phase{1}. This is consistent with the values of the capacitances $\Cinfpn{\aC}=\infty$ and $\Con{\aC} = \infty$ extracted from the circuits of \cref{fig:SC_int_ph1abcd} which make the second and  third terms of \cref{eqn:V2n_Bode_OTA} due to the OTA zero. The variance of the noise voltage sampled on capacitor $\aC$ at the end of \phase{1} is then simply given by
\begin{equation}\label{eqn:SC_int_V_aC_ph1}
  \Varp{V}{\aC}{1} = \kT \cdot \left[\frac{1}{\aC} + 0 - 0 \right] = \frac{\kT}{\aC}.
\end{equation}
The corresponding noise charge sampled at the end of \phase{1} and transferred to capacitor $C$ during \phase{2} is given by
\begin{equation}\label{eqn:SC_int_Q_ph1}
  \Varp{Q}{}{1} = \kT \cdot \aC.
\end{equation}

At the end of \phase{2}, because of the noise coming from the switches S\sub{1} and S\sub{2} and from the OTA, not all the charge sampled on $\aC$ are transferred to $C$ and some will remain on $\aC$ after the opening of switch S\sub{2} leading to a random charge ``deficit'' on $C$. The latter can be modelled by the charge $\left.Q_n\right|_{\Phi_2}$ which is equal to the charge sampled on $\aC$ at the end of \phase{2}. The variance of this charge can be calculated from the variance of the voltage across $\aC$ during \phase{2} which can be estimated by applying the extended Bode theorem \cref{eqn:V2n_Bode_OTA} to capacitor \aC{} during \phase{2}. Capacitances \Cinfn{\aC}, \Cinfpn{\aC} and \Con{\aC} together with the feedback gain \hfb{} during \phase{2} are calculated with the help of the schematics shown in \cref{fig:SC_int_ph2abcd}. The noise voltage variance across \aC{} during \phase{2} is derived in details in Appendix~\ref{sec:SC_int_V2naCp2}. Assuming that $\aC \ll C$ and $\Cin \ll C$ (i.e. $\a \ll 1$ and $\ain \ll 1$), leads to
\begin{equation}\label{eqn:SC_int_V2naCa2}
  \Varp{V}{\aC}{2} \cong \frac{\kT}{\aC} \cdot \frac{\aL+\ain+\gamma \cdot \a}{\aL+\a+\ain},
\end{equation}
where $\ain \triangleq C_{in}/C$ and $\aL \triangleq C_L/C$. The corresponding variance of the noise charge sampled on $\aC$ at the end of \phase{2} is then given by
\begin{equation}\label{eqn:SC_int_Q_C_ph2}
  \Varp{Q}{}{2} = \kT \cdot \aC \cdot \frac{\aL+\ain+\gamma \cdot \a}{\aL+\a+\ain}.
\end{equation}

Both noise charges \Varp{Q}{}{1} and \Varp{Q}{}{2} add to the charge already held on capacitor $C$. This noise charge injection on $C$ can be modeled by the charge injector defined by \cref{eqn:Q2n} and given by
\begin{equation}\label{eqn:SC_int_Q2n}
    \Var{Q}{} = \kT \cdot \aC \cdot \left(1+\frac{\aL+\ain+\gamma \cdot \a}{\aL+\a+\ain}\right).
\end{equation}

To evaluate the output noise voltage variance, let's first consider \Qn{C}{n} as the variance of the noise charge stored on capacitor $C$ at the end of the $n^{th}$ switching period. The noise charge variances $Q_n^2$ calculated above will add to \Qn{C}{n} so that the total noise charge held on capacitor $C$ at the end of the $(n+1)^{th}$ switching period is given by
\begin{equation}\label{eqn:SC_int_ch_eq}
  \Qn{C}{n+1} = \Qn{C}{n} + Q_n^2.
\end{equation}
\Cref{eqn:SC_int_ch_eq} shows that the sampled noise charge at the $n^{th}$ period is not shared with any other capacitor and is entirely added at the $(n+1)^{th}$ period to the noise charge already held on $C$. Therefore the resulting sampled noise charge is the noise generated during a single period multiplied by the number of switching periods $n$. The variance of the total noise charge cumulated on $C$ after $n$ switching periods, assuming $\Qn{C}{n=0}=0$, is then simply given by $n \cdot Q_n^2$ resulting in
\begin{equation}\label{eqn:SC_int_Q_C_diverg}
  \Qn{C}{n} = n \cdot \kT \cdot \aC \cdot \left(1+\frac{\aL+\ain+\gamma \cdot \a}{\aL+\a+\ain}\right).
\end{equation}
The corresponding variance of the noise voltage at the OTA output after $n$ switching periods is then given by
\begin{equation}\label{eqn:SC_int_V_C_diverg}
  \Varn{V}{,sampled}{n}= n \cdot \frac{\kT \cdot \a}{C} \cdot \left(1+\frac{\aL+\ain+\gamma \cdot \a}{\aL+\a+\ain}\right).
\end{equation}

As expected for an integrator, \cref{eqn:SC_int_V_C_diverg} shows that the output noise voltage variance increases proportionally with the switching periods $n$, which is typical of a Wiener process \cite{bib:papoulis:book:1981}.

\Cref{eqn:SC_int_V_C_diverg} does not account for the direct noise at the output of the OTA which would be sampled by the next stage at the end of \phase{1}. The latter adds to the sampled noise given by \cref{eqn:SC_int_V_C_diverg}. It can be calculated by applying the extended Bode theorem \cref{eqn:V2n_Bode_OTA} applied on the load capacitor $\CL$ during the readout \phase{1}. The derivation is detailed in Appendix~\ref{sec:SC_int_direct}. Assuming again that $\alpha \ll 1$ and $\ain \ll 1$, it results in
\begin{equation}\label{eqn:SC_int_direct}
  \Varp{V}{,direct}{1} \cong \frac{\gamma \cdot \kT }{\CL+\Cin}.
\end{equation}

Finally the total output noise voltage variance during the readout \phase{1} is given by
\begin{equation}\label{eqn:SC_int_total}
  \begin{split}
    \Varn{V}{out}{n} =& n \cdot \frac{\kT \cdot \a}{C} \cdot \left(1+\frac{\aL+\ain+\gamma \cdot \a}{\aL+\a+\ain}\right)\\
    &+ \frac{\gamma \cdot \kT }{\CL+\Cin}.
  \end{split}
\end{equation}

\begin{figure}[!t]
  \centering
  \begin{subfigure}[t]{1\columnwidth}
    \centering
    \includegraphics[width=0.7\columnwidth]{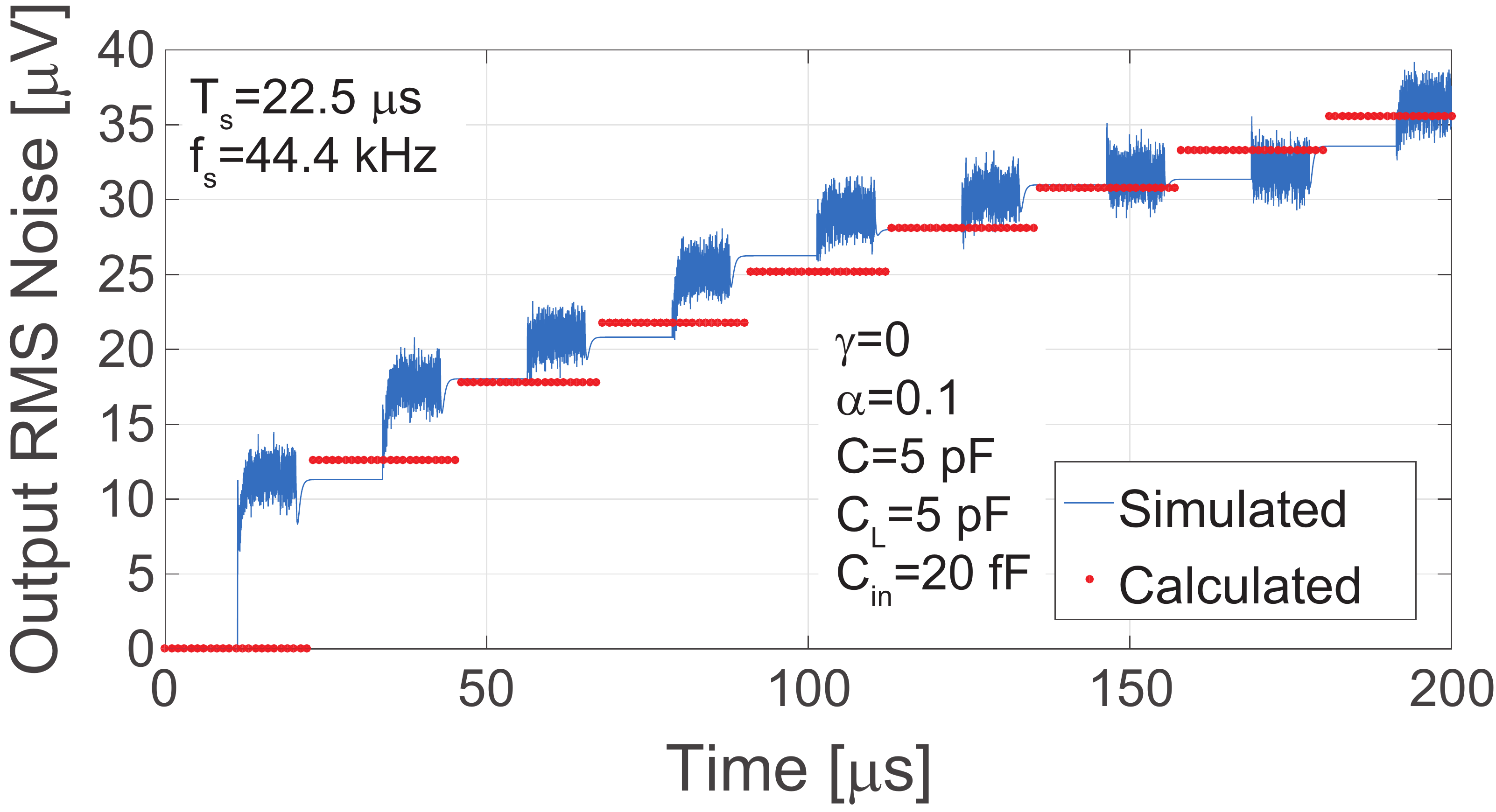}
    \caption{\centering $\gamma=0$ (i.e. ignoring the noise coming from the OTA).}
    \label{fig:SC_int_trans_gamma0}
  \end{subfigure}%
  \\
  \begin{subfigure}[t]{1\columnwidth}
  \centering
  \includegraphics[width=0.7\columnwidth]{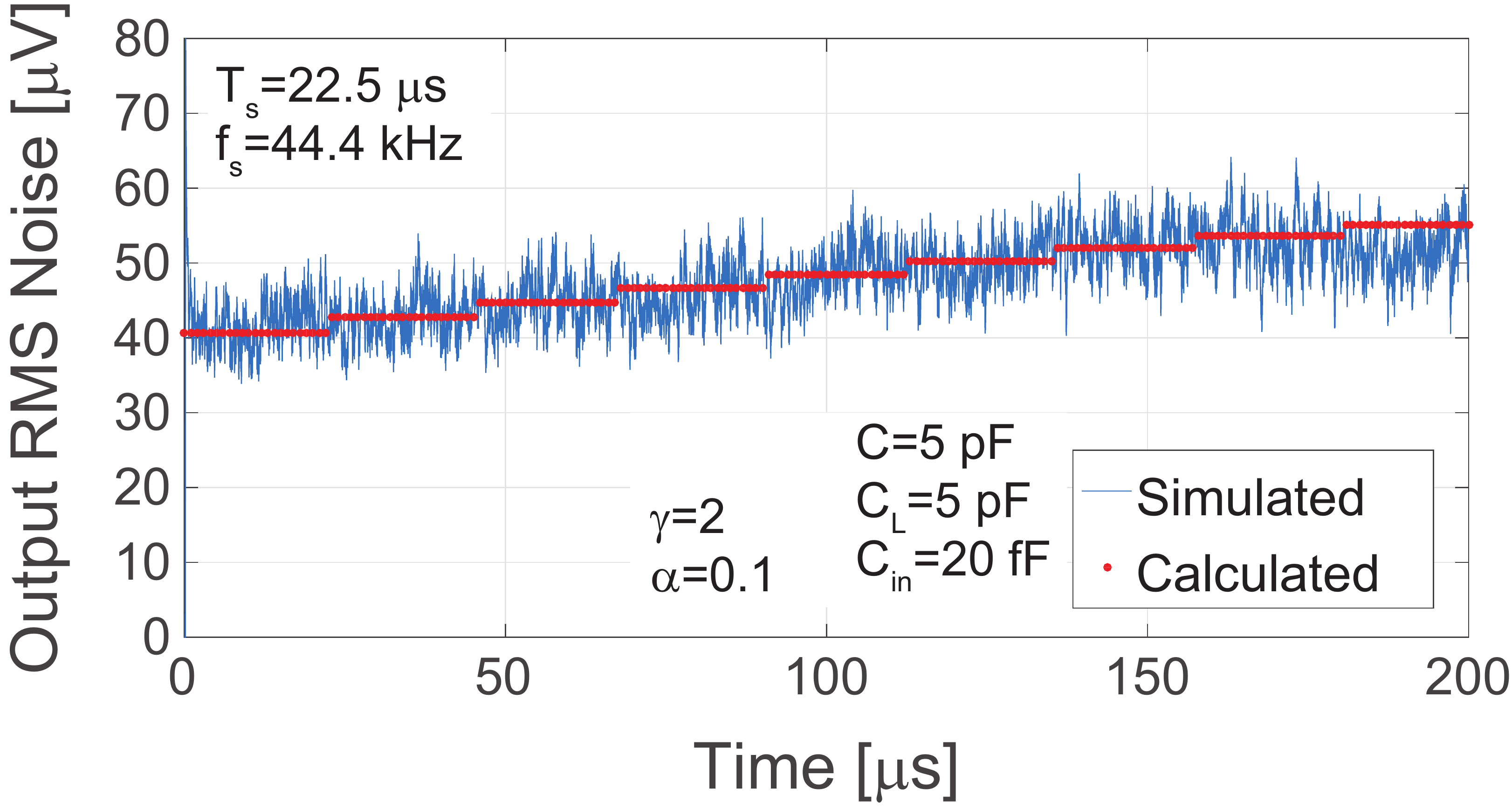}
  \caption{\centering $\gamma=2$.}
  \label{fig:SC_int_trans_gamma2}
  \end{subfigure}
  \caption{OTA-based SC integrator output RMS noise voltage simulated by transient noise and compared to the voltage calculated from \cref{eqn:SC_int_total} for a sampling frequency $f_s=44.4\,kHz$, $\alpha=0.1$, $C=\CL=5\,pF$ and $\Cin=20\,fF$.}
  \label{fig:SC_int_trans}
\end{figure}

\Cref{eqn:SC_int_total} has been verified using transient noise simulations. \Cref{fig:SC_int_trans_gamma0} shows the transient simulation of the output RMS noise voltage in the case of a noiseless OTA (i.e. $\gamma=0$) compared to the calculated noise voltage using \eqref{eqn:SC_int_total}. Note that the calculated noise corresponds to the simulated noise for the readout \phase{1} during which the output voltage remains constant since the OTA does not generate any direct noise during this phase when $\gamma$ is set to zero.

\Cref{fig:SC_int_trans_gamma2} presents the simulation for an OTA with a noise excess factor $\gamma=2$ generating the additional direct noise during the readout \phase{1} having an RMS value of $40.7\;\mu V_{rms}$. Both simulations show the increase of the output RMS noise voltage following a $\sqrt{n}$ law. \Cref{fig:SC_int_trans} shows that the simulations precisely match the RMS output noise voltage calculated from \cref{eqn:SC_int_total}. The excellent fit between the calculated and simulated noise confirms the effectiveness of the proposed noise calculation method.

\subsection{Active First-order LP Filter Based on OTA}

\begin{figure}[!t]
  \centering
  \begin{subfigure}[t]{1\columnwidth}
    \centering
    \includegraphics[width=0.7\columnwidth]{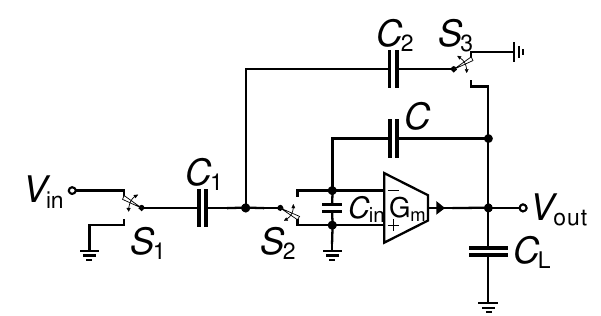}
    \caption{\centering \Phase{1}}
    \label{fig:SC_LP_Act_ph1}
  \end{subfigure}%
  \\
  \begin{subfigure}[t]{1\columnwidth}
    \centering
    \includegraphics[width=0.7\columnwidth]{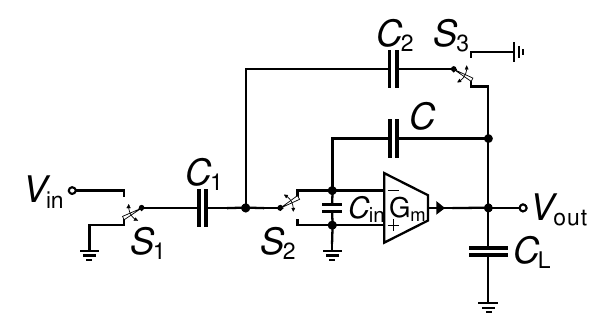}
    \caption{\centering \Phase{2}}
    \label{fig:SC_LP_Act_ph2}
  \end{subfigure}
  \caption{OTA-based SC first-order low-pass filter.}
  \label{fig:SC_LP_Act}
\end{figure}

The active SC LP filter is implemented using an OTA as shown in \cref{fig:SC_LP_Act}. The input signal is sampled by capacitor $C_1$ during \phase{1} while capacitor $C_2$ is discharged and capacitor $C$ holds the charge transferred during the previous period. The charge sampled on $C_1$ at the end of \phase{1} is then transferred to $C_2$ and $C$ during \phase{2}. The transfer function of this filter in the $z$ domain is given by
\begin{equation}\label{eqn:SC_LP_Act_transfer_function}
  H(z)=\frac{\alpha_1 \cdot z^{-1}}{1+\alpha_2-z^{-1}},
\end{equation}
where $\alpha_1=C_1/C$ and $\alpha_2=C_2/C$. Usually $\alpha_1$ and $\alpha_2$ are much smaller than unity to set the cut-off frequency at a much smaller value than the sampling frequency. In order to further simplify the calculations, the ratios can be assumed equal and much smaller than unity, $\alpha_1 = \alpha_2 = \alpha \ll 1$. For frequencies much smaller than the sampling frequency, this circuit operates as first-order LP filter with a cutoff frequency given by
\begin{equation}\label{eqn:SC_LP_Act_cutoff}
  f_c=\frac{\alpha}{2\pi} \cdot f_s,
\end{equation}
where $f_s$ is the sampling frequency.

In this example, the readout of the output voltage is performed at the end of \phase{1}. Therefore, the output noise is composed of the noise sampled on the integrating capacitor $C$ at the end of \phase{2} of the previous switching period and held over to \phase{1} of the next switching period, and the direct output noise sampled on $\CL$ at the end of \phase{1} and originating from the OTA during this readout \phase{1}. Hence, the noise calculation starts with the evaluation of the total noise sampled on the integrating capacitor $C$ and coming from \phase{1} and \phase{2}, and follows with the estimation of the direct noise sampled at the end of \phase{1}.

\begin{figure*}[!t]
  \centering
  \begin{subfigure}[t]{.24\textwidth}
    \centering
    \includegraphics[scale=.6]{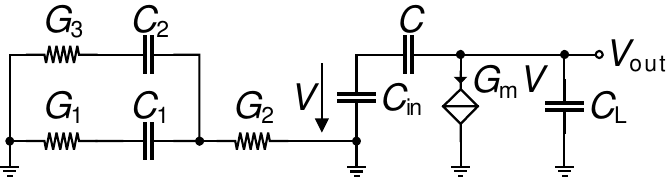}
    \caption{\centering Equivalent circuit of the first-order OTA-based SC LP filter during \phase{1}.}
    \label{fig:SC_LP_Act_ph1a}
  \end{subfigure}
  \begin{subfigure}[t]{.24\textwidth}
    \centering
    \includegraphics[scale=.6]{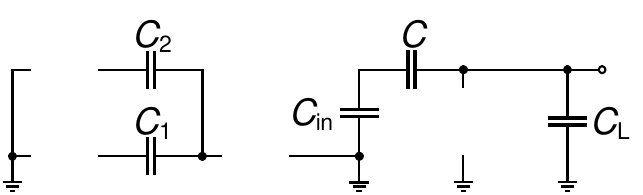}
    \caption{\centering $\Cinfn{C_1}=C_1$\\
    $\Cinfn{C_2}=C_2$\\
    $\Cinfn{\CL}=\CL $.}
    \label{fig:SC_LP_Act_ph1b}
  \end{subfigure}
  \begin{subfigure}[t]{.24\textwidth}
    \centering
    \includegraphics[scale=.6]{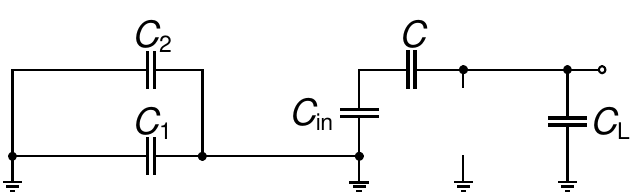}
    \caption{\centering $\Cinfpn{C_1}=\Cinfpn{C_2}=\infty$\\
    $\Cinfpn{\CL}=\CL $.}
    \label{fig:SC_LP_Act_ph1c}
  \end{subfigure}
  \begin{subfigure}[t]{.24\textwidth}
    \centering
    \includegraphics[scale=.6]{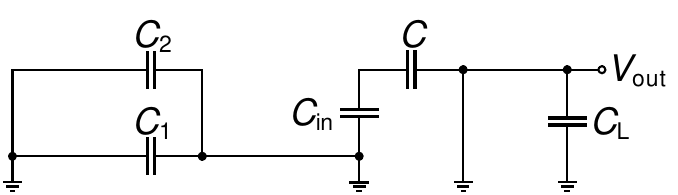}
    \caption{\centering $\Con{C_1}=\Con{C_2}=\Con{\CL}=\infty$}
    \label{fig:SC_LP_Act_ph1d}
  \end{subfigure}
  \caption{Equivalent circuit schematics for the calculation of the noise voltage variances of the first-order LP filter in \phase{1}.}
  \label{fig:SC_LP_Act_ph1abcd}
\end{figure*}

\begin{figure*}[!hbt]
  \centering
  \begin{subfigure}[t]{.24\textwidth}
    \centering
    \includegraphics[scale=.6]{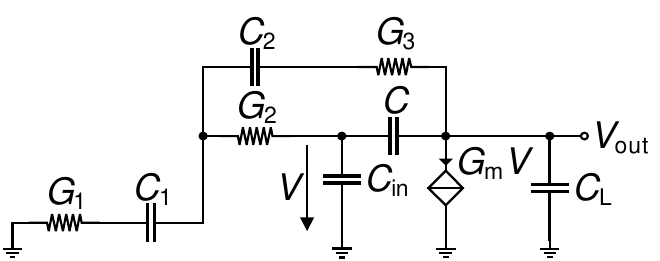}
    \caption{\centering  Equivalent circuit of the first-order OTA-based SC LP filter during  \phase{2}.}
    \label{fig:SC_LP_Act_ph2a}
  \end{subfigure}
  \begin{subfigure}[t]{.24\textwidth}
    \centering
    \includegraphics[scale=.6]{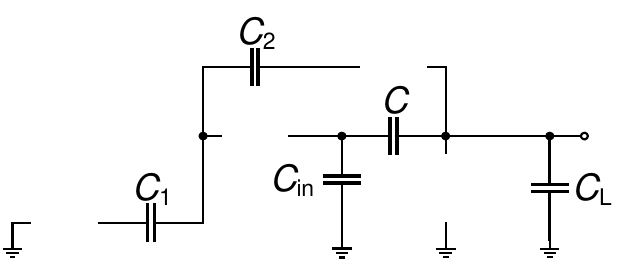}
    \caption{\centering $\Cinfn{C}=C$}
    \label{fig:SC_LP_Act_ph2b}
  \end{subfigure}
  \begin{subfigure}[t]{.24\textwidth}
    \centering
    \includegraphics[scale=.6]{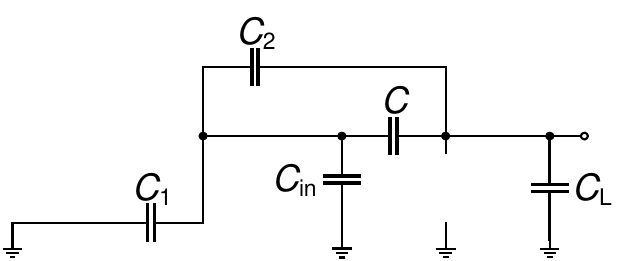}
    \caption{\centering $\Cinfpn{C}=C+C_2+\frac{C_1 C_L}{C_1+C_L}$}
    \label{fig:SC_LP_Act_ph2c}
  \end{subfigure}
  \begin{subfigure}[t]{.24\textwidth}
    \centering
    \includegraphics[scale=.6]{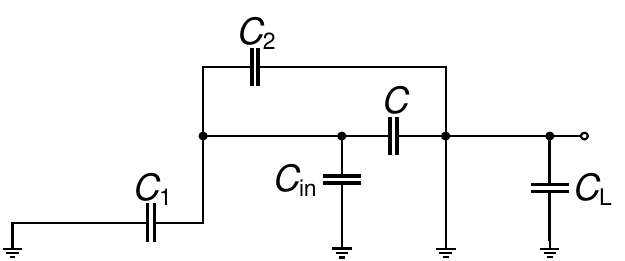}
    \caption{\centering $\Con{C}=C+C_1+C_2$}
    \label{fig:SC_LP_Act_ph2d}
  \end{subfigure}
  \caption{Equivalent circuit schematics for the calculation of the noise voltage variances of the first-order LP filter in \phase{2}.}
  \label{fig:SC_LP_Act_ph2abcd}
\end{figure*}

At the end of \phase{1}, capacitors $C_1$ and $C_2$ sample a noise charge generated by switches S\sub{1}, S\sub{2} and S\sub{3}. The sum of these noise charges is then injected into the virtual ground and transferred to the feedback capacitors $C$ and $C_2$ during the charge transfer \phase{2}. The noise voltage variances across capacitors $C_1$ and $C_2$ during \phase{1} can be calculated applying the extended Bode theorem \cref{eqn:V2n_Bode_OTA} with the use of the schematics shown in \cref{fig:SC_LP_Act_ph1abcd} resulting in
\begin{subequations}
  \begin{align}
    \Varp{V}{C_1}{1} &= \kT \cdot \left[\frac{1}{C_1} + 0 - 0 \right] = \frac{\kT}{C_1} = \frac{\kT}{\aC},\label{eqn:SC_LP_Act_V_C1_ph1}\\
    \Varp{V}{C_2}{1} &= \kT \cdot \left[\frac{1}{C_2} + 0 - 0 \right] = \frac{\kT}{C_2} = \frac{\kT}{\aC}.\label{eqn:SC_LP_Act_V_C2_ph1}
  \end{align}
\end{subequations}

The corresponding noise charge sampled during \phase{1} and injected into the virtual ground during \phase{2} is given by
\begin{equation}\label{eqn:SC_LP_Act_Q_ph1}
  \Varp{Q}{}{1} = C_1^2 \cdot \Varp{V}{C_1}{1} + C_2^2 \cdot \Varp{V}{C_2}{1} = \kT \cdot 2 \aC.
\end{equation}
This noise charge $\Qnp{}{1}$ is then shared by the parallel capacitors $C$ and $C_2$ during \phase{2}. Consequently, only a fraction $C/(C+C_2)=1/(1+\alpha)$ of this charge will remain on capacitor $C$ after the end of \phase{2} when capacitor $C_2$ is disconnected from $C$. Hence, the variance of the noise charge generated during \phase{1} and remaining on $C$ after the end of \phase{2} is given by
\begin{equation}\label{eqn:SC_LP_Act_Q_C_ph1}
  \Varp{Q}{C}{1} = \left(\frac{C}{C+C_2}\right)^2 \cdot \Varp{Q}{}{1} = \kT \cdot C \cdot \bswip{1},
\end{equation}
where $\bswip{1} = 2 \a/(1+\a)^2$.

The noise charge that is injected into the virtual ground at the end of \phase{2} actually corresponds to the noise charge $\Qnp{C}{2}$ that is sampled on the integrating capacitor $C$ when capacitors $C_1$ and $C_2$ are disconnected from the virtual ground and the output node, respectively. The variance of this noise charge is calculated from the variance of the noise voltage across capacitor $C$ at the end of \phase{2} due to the noise of the switches and OTA generated during \phase{2}. The latter is derived using the extended Bode theorem \cref{eqn:V2n_Bode_OTA} in Appendix~\ref{sec:SC_LP_Act_V2nCp2}. It can be written separating the contributions of the switches and OTA as
\begin{equation}\label{eqn:SC_LP_Act_V2nCp2}
  \Varp{V}{C}{2} = \frac{\kT}{C} \cdot \left(\gamma \cdot \botap{2} + \bswip{2}\right),
\end{equation}
where \botap{2} and \bswip{2} are given by \cref{eqn:SC_LP_Act_beta_sw2} and \cref{eqn:SC_LP_Act_beta_ota}, respectively. For $\a \ll 1$ and $\ain \triangleq \Cin/C \ll 1$, \botap{2} and \bswip{2} reduce to
\begin{subequations}
  \begin{align}
    \botap{2} &\cong \frac{(\a+\ain)^2}{\aL+\a+\ain},\label{eqn:SC_LP_Act_beta_ota_a}\\
    \bswip{2} &\cong \a \cdot \left(1+\frac{\aL^2}{(\aL+\ain)(\aL+\a+\ain)}\right),\label{eqn:SC_LP_Act_beta_sw_a}
  \end{align}
\end{subequations}
with $\aL \triangleq \CL/C$. The corresponding noise charge variance sampled and remaining on $C$ at the end of \phase{2} (ignoring the noise charge coming from \phase{1}) is expressed as
\begin{equation}\label{eqn:SC_LP_Act_V_C_ph2b}
  \Varp{Q}{C}{2} = \kT \cdot C \cdot \left(\gamma \cdot \botap{2} + \bswip{2}\right).
\end{equation}

The variance of the total noise charge injected into capacitor $C$ at the end of \phase{2} due to the noise generated during phases \ph{1} and \ph{2} is then given by
\begin{equation}\label{eqn:SC_LP_Act_Q2nC}
  \begin{split}
    \Var{Q}{} &= \Varp{Q}{C}{1} + \Varp{Q}{C}{2}\\
              &= \kT \cdot C \cdot \left(\gamma \cdot \bota + \bswi\right).
  \end{split}
\end{equation}
where
\begin{subequations}
  \begin{align}
    \bota &= \botap{2},\label{eqn:SC_LP_Act_beta_ota_tot}\\
    \bswi &= \bswip{1} + \bswip{2}.\label{eqn:SC_LP_Act_beta_switch_tot}
  \end{align}
\end{subequations}

Similarly to the passive first-order SC filter discussed above, capacitor $C$ is not reset and will cumulate part of this injected noise charge. During the $n+1$ switching period, the noise charge $Q_n$, which variance is calculated above, will add to the noise charge $Q_{nC}(n)$ already present on capacitor $C$ at the end of the switching period $n$. The charge $Q_{nC}(n)$ is shared between capacitors $C_2$ and $C$ during \phase{2} of the switching period $n+1$, and when capacitor $C_2$ is disconnected from $C$ at the end of \phase{2}, only a fraction $C/(C+C_2)=1/(1+\alpha)$ of this charge will remain on $C$. The variance of the total noise charge sampled on $C$ at the end of the $(n+1)^{th}$ switching period can therefore be expressed as
\begin{equation}\label{eqn:SC_LP_Act_ch_eq}
  \Varn{Q}{C}{n+1} =  \frac{\Varn{Q}{C}{n}}{(1+\a)^2} + \Var{Q}{}.
\end{equation}
\Cref{eqn:SC_LP_Act_ch_eq} corresponds to a recurrence equation similar to the example of the passive SC LP filter. The variance of the noise charge held on capacitor $C$ at the $n^{th}$ switching period can hence be expressed as
\begin{equation}\label{eqn:SC_LP_Act_conv}
  \Varn{Q}{C}{n} = \Var{Q}{C\infty} \cdot \left[1- \left(\frac{1}{1+\a}\right)^{2n}\right].
\end{equation}
where
\begin{equation}\label{eqn:SC_LP_Qinf}
  \Var{Q}{C\infty} = \Var{Q}{} \cdot \frac{(1+\a)^2}{\a(2+\a)} \cong \frac{\Var{Q}{}}{2\a}.
\end{equation}

After several switching periods the right term of \cref{eqn:SC_LP_Act_conv} tends to unity. Hence, the variance of the noise voltage sampled on capacitor $C$ tends to the value
\begin{equation}\label{eqn:SC_LP_Act_Vn_sampled}
  \Var{V}{,sampled} = \frac{\kT}{C} \cdot \left(\gamma \cdot \tota + \tswi\right),
\end{equation}
where $\tota = \totap{2}$ and $\tswi=\tswip{1}+\tswip{2}$ with \totap{2}, \tswip{1} and \tswip{2} given by
\begin{subequations}
  \begin{align}
    \begin{split}
      \totap{2} &= \frac{(1+\a)^2}{\a(2+\a)} \cdot \botap{2} \cong \frac{\botap{2}}{2\a} =\\
                &= \frac{(\a+\ain)^2}{2\a(\aL+\a+\ain)},\label{eqn:SC_LP_Act_beta_ota2}
    \end{split}\\
    \tswip{1} &= \frac{(1+\a)^2}{\a(2+\a)} \cdot \bswip{1} \cong \frac{\bswip{1}}{2\a} = 1,\label{eqn:CS_LP_Act_beta_sw1_tot}\\
    \begin{split}
      \tswip{2} &= \frac{(1+\a)^2}{\a(2+\a)} \cdot \bswip{2} \cong \frac{\bswip{2}}{2\a} =\\
                &= \frac{1}{2} \cdot \left[1+\frac{\aL^2}{(\aL+\ain)(\aL+\a+\ain)}\right].\label{eqn:SC_LP_Act_beta_sw2_tot}
    \end{split}
  \end{align}
\end{subequations}

Since the output voltage is read (or sampled) at the end of \phase{1}, the direct noise that appears at the OTA output during this readout \phase{1} adds to the sampled noise \cref{eqn:SC_LP_Act_Vn_sampled}. The variance of this direct output noise voltage can also be calculated using the extended Bode theorem \cref{eqn:V2n_Bode_OTA}. The derivation of capacitances $C_{\infty}$, $C_{\infty}'$ and $C_{0}$  can be done with the help of the schematics of \cref{fig:SC_LP_Act_ph1abcd} but applied to the load capacitor $C_L$. The feedback gain $\hfb$ during \phase{1} which is also required in \cref{eqn:V2n_Bode_OTA}, is $h_{fb}=1/(1+\ain) \cong 1$. The variance of the output voltage during \phase{1} therefore reduces to
\begin{equation}\label{eqn:SC_LP_Act_Vn_direct}
  \Var{V}{,direct} = \frac{\gamma \cdot \kT}{\CL+\Cin} = \frac{\kT}{C} \cdot \gamma \cdot \tdir,
\end{equation}
where $\tdir \triangleq 1/(\aL+\ain)$.

\begin{figure}[!t]
  \centering
  \begin{subfigure}[t]{1\columnwidth}
    \centering
    \includegraphics[width=0.8\columnwidth]{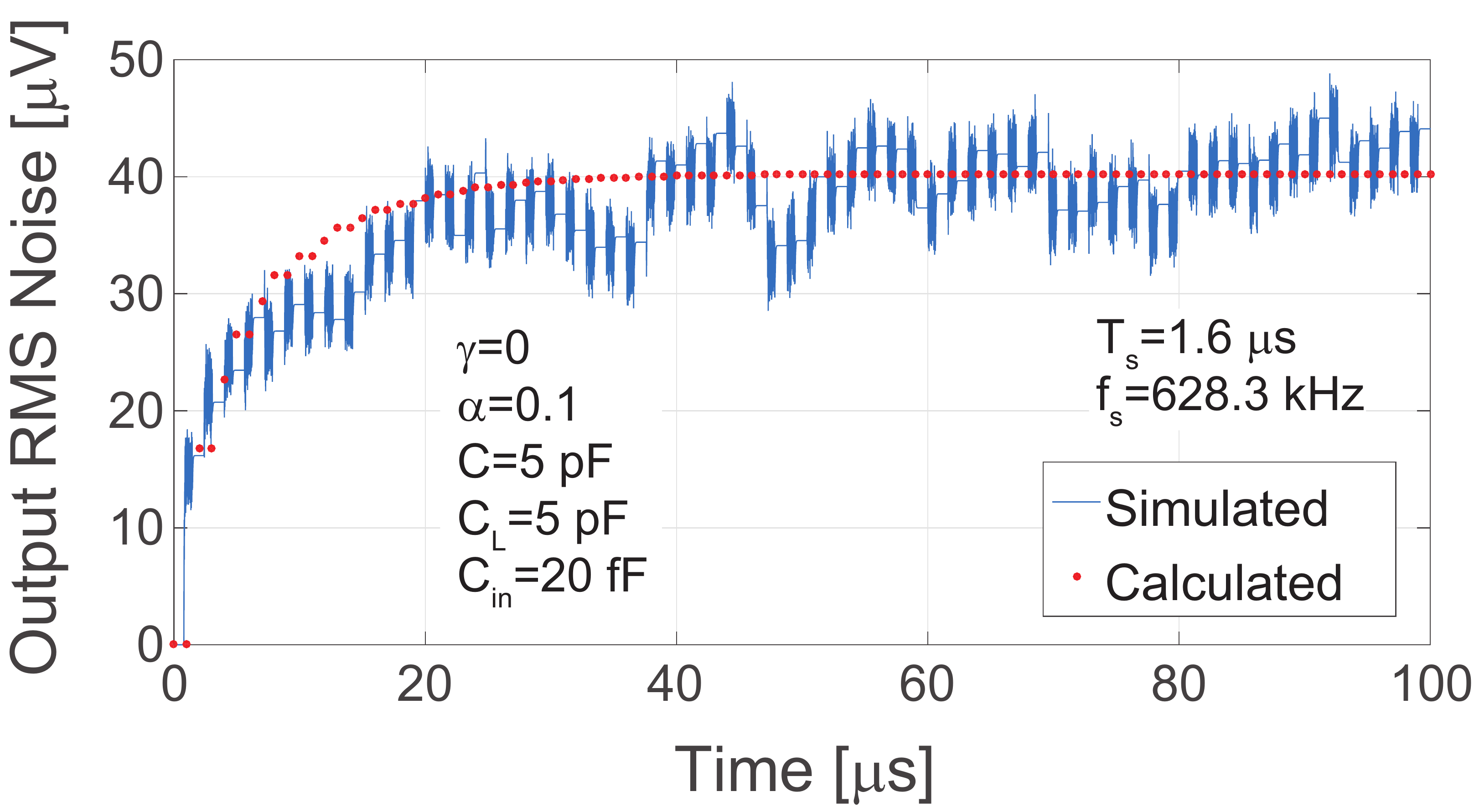}
	\caption{\centering$\alpha=0.1$, $C=C_L=5\,pF$, $\Cin=20\,fF$ and $\gamma=0$ (i.e. ignoring the noise coming from the OTA) resulting in $V_{n,out}=40\,\mu V_{rms}$ at $T=300\,K$.}
	\label{fig:SC_LP_act_trans_gamma0}
  \end{subfigure}%
  \\
  \begin{subfigure}[t]{1\columnwidth}
    \centering
	\includegraphics[width=0.8\columnwidth]{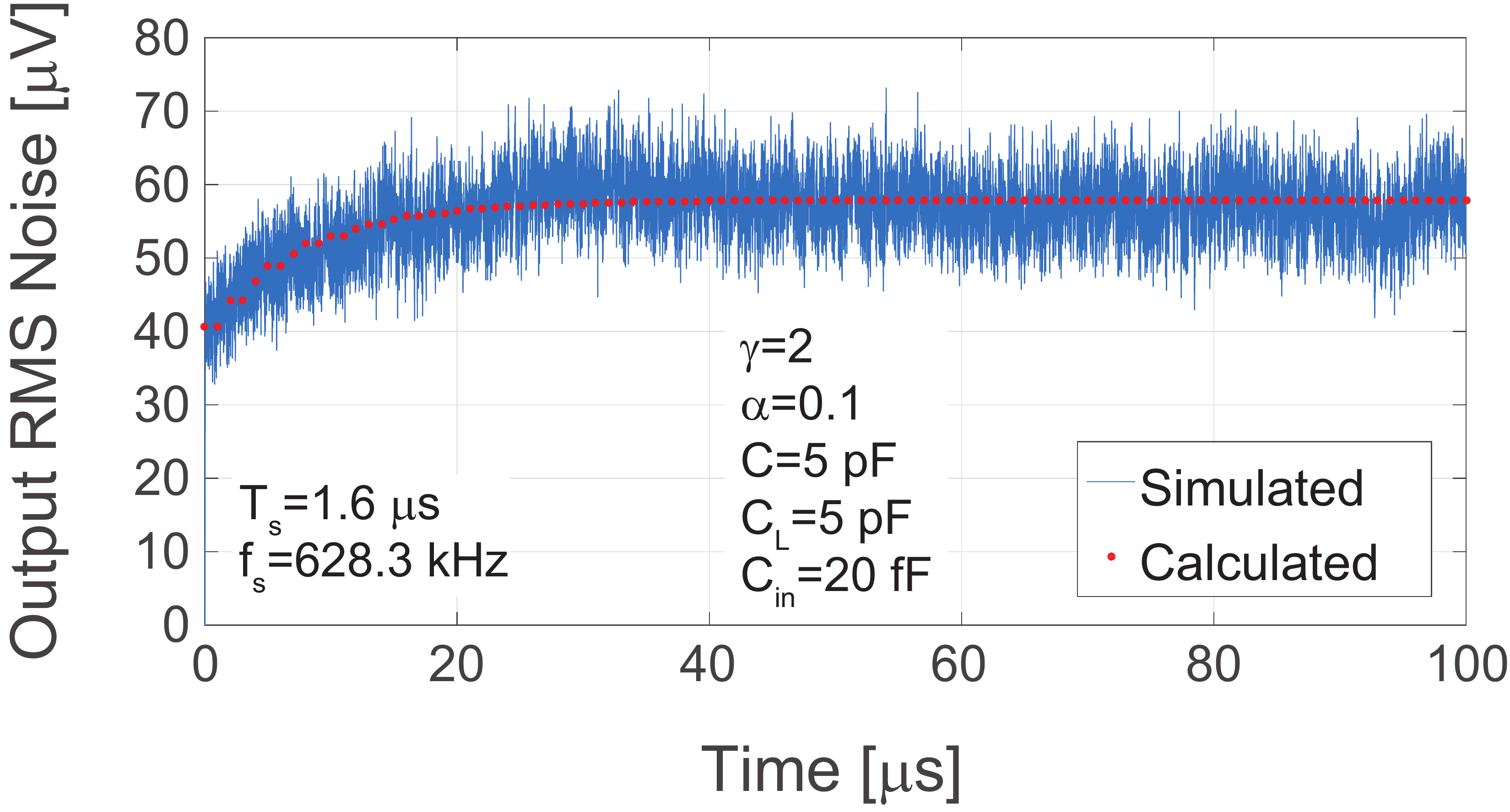}
	\caption{\centering$\alpha=0.1$, $C=C_L=5\,pF$, $\Cin=20\,fF$ and $\gamma=2$ resulting in $V_{n,out}=58\,\mu V_{rms}$ at $T=300\,K$.}
	\label{fig:SC_LP_act_trans_gamma2}
  \end{subfigure}
  \\
  \begin{subfigure}[t]{1\columnwidth}
    \centering
	\includegraphics[width=0.8\columnwidth]{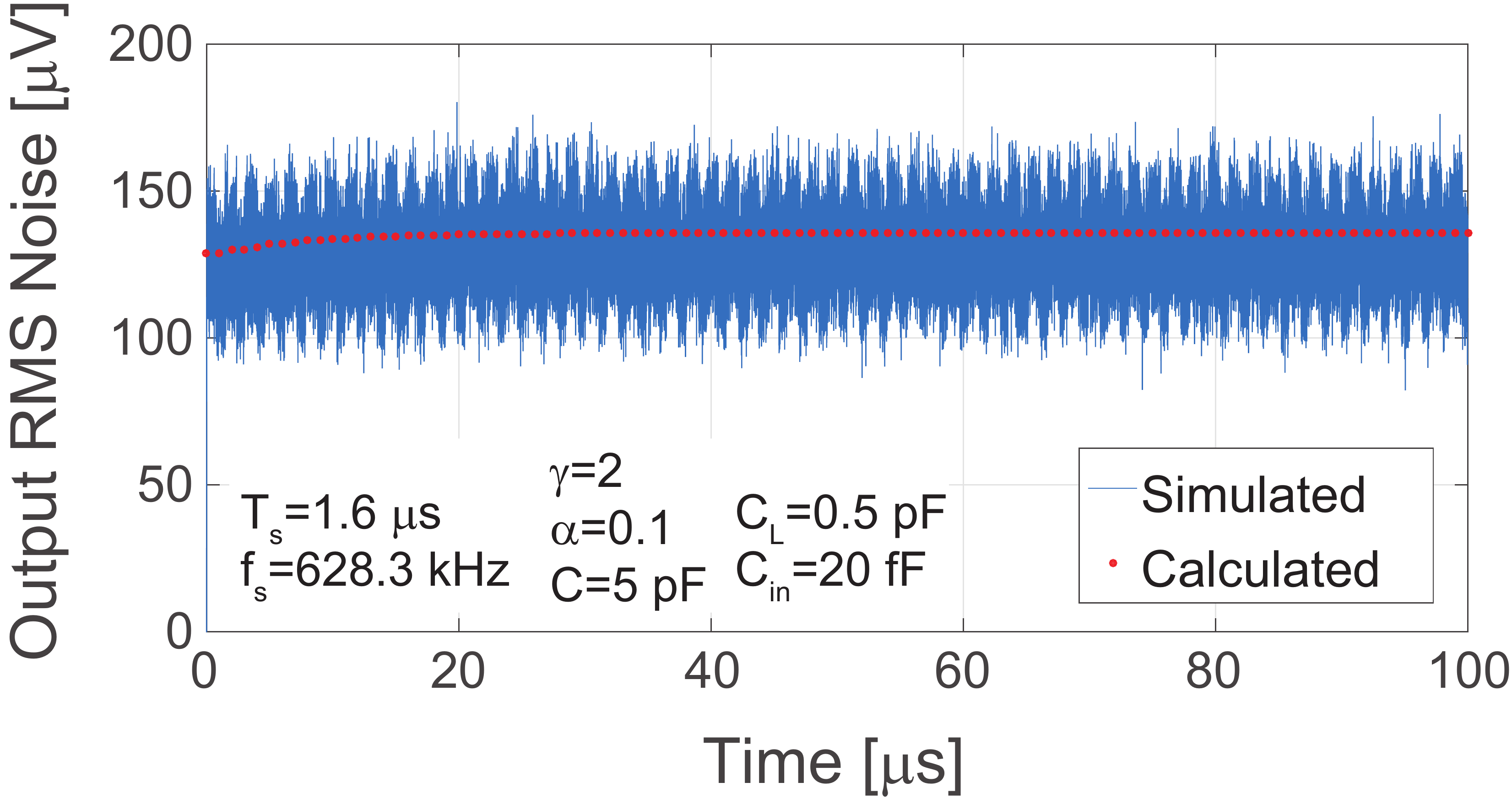}
	\caption{\centering$\alpha=0.1$, $C=5pF$, $C_L=0.5pF$, $\Cin=20\,fF$ and $\gamma=2$ resulting in $V_{n,out}=133\,\mu V_{rms}$ at $T=300\,K$.}
	\label{fig:SC_LP_act_trans_gamma2b}
  \end{subfigure}
  \caption{OTA-based SC first-order LP filter output RMS noise voltage simulated by transient noise and compared to the voltage calculated from \cref{eqn:SC_LP_Act_Vn_total} for different values of $\alpha$, $C$, $C_L$ and $\gamma$.}
  \label{fig:SC_LP_act_trans}
\end{figure}

\begin{figure}[!t]
  \centering
  \includegraphics[width=1.0\columnwidth]{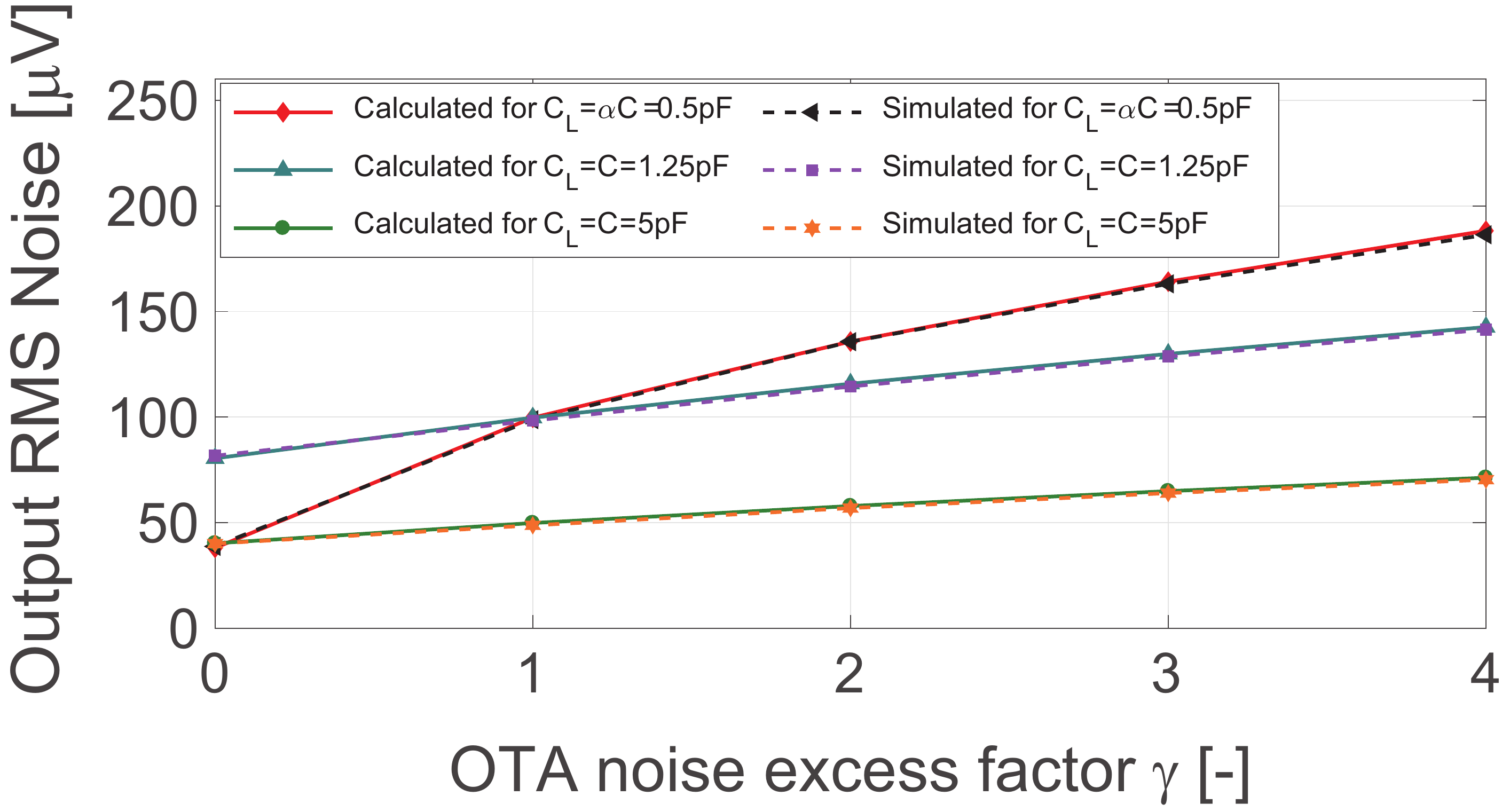}
  \caption{Simulated and calculated output noise RMS voltage versus OTA noise excess factor for the SC OTA-based LP filter with different capacitors.}
  \label{fig:SC_LP_act_gamma}
\end{figure}

The total output noise voltage variance is then simply given by summing \cref{eqn:SC_LP_Act_Vn_sampled} and \cref{eqn:SC_LP_Act_Vn_direct} resulting in
\begin{equation}\label{eqn:SC_LP_Act_Vn_total}
  \begin{split}
    \Var{V}{,out} &= \Var{V}{,direct} +\Var{V}{,sampled} =\\
    &= \frac{\kT}{C} \cdot \left[\gamma \cdot \left(\tota + \tdir\right) + \tswi\right].
  \end{split}
\end{equation}
In case $\CL$ can be assumed to be much larger than $\aC$ and $\Cin$ (i.e. $\aL \gg \a$ and $\aL \gg \ain$), then $\theta_{sw2}\cong1$ and the noise contribution of the OTA to the sampled noise can be neglected (i.e. $\theta_{ota} \ll 1$) and \cref{eqn:SC_LP_Act_Vn_total} reduces to
\begin{equation}\label{eqn:SC_LP_Act_Vn_total_approx}
  \Var{V}{,out} \cong \frac{\gamma \cdot \kT }{\CL} + \frac{2\kT}{C}.
\end{equation}

\begin{figure*}[!t]
\normalsize
\setcounter{mytempeqncnt}{\value{equation}}
\setcounter{equation}{53}
\begin{equation}\label{eqn:SC_LP_Act_beta_ota}
  \botap{2} = \frac{(\a+\ain)^2}{(1+\a)(\aL(1+2\a+\ain)+(1+\a)(\a+\ain))},
  \quad \text{where} \;
  \ain \triangleq \frac{\Cin}{C},
  \aL \triangleq \frac{\CL}{C}.
\end{equation}
\begin{equation}\label{eqn:SC_LP_Act_beta_sw2}
  \bswip{2} = \frac{\a((\aL+\ain)(\aL+\a+\ain)+\aL^2)}{(\aL(1+\ain)+\ain)(\aL(1+2\a+\ain)+(1+\a)(\a+\ain))},
\end{equation}
\setcounter{equation}{\value{mytempeqncnt}}
\hrulefill
\vspace*{4pt}
\end{figure*}

In order to validate the above results, the circuit is simulated for different values of capacitors and noise excess factor. \Cref{fig:SC_LP_act_trans_gamma0} shows the transient behavior of the output RMS noise for $\alpha=0.1$, $C=C_L=5\,pF$, $\Cin=10\,fF$ and a noiseless OTA ($\gamma=0$) resulting in an output RMS noise voltage $V_{n,out}=40.2\,\mu V_{rms}$ at $T=300\,K$. Note that the simulated noise during \phase{1} is constant because the direct noise is not present in this phase when the OTA is considered noiseless. The simulations in \cref{fig:SC_LP_act_trans_gamma2,fig:SC_LP_act_trans_gamma2b} take into account the noise originated from the OTA with $\gamma=2$ and different values of the load capacitance. The simulations show a direct noise but the value is not the same during both phases because the continuous-time circuit for each phase is different. The expression in \cref{eqn:SC_LP_Act_Vn_total} is the noise calculated during \phase{1} and it is slightly higher than the noise during \phase{2}. This can be specially observed in \cref{fig:SC_LP_act_trans_gamma2b} due to the small load capacitor which increases the direct noise during \phase{2} compared to $\Phi_1$.

The excellent match between the transient noise simulations and the results calculated from \cref{eqn:SC_LP_Act_Vn_total} and presented in \cref{fig:SC_LP_act_trans} demonstrates the efficiency of the proposed noise estimation method in accurately predicting the value of the noise variance as well as how exactly the output noise converges to its steady-state value.

In order to have a broader comparison, \cref{fig:SC_LP_act_gamma} shows the calculated and simulated output RMS noise voltage versus the OTA noise excess factor $\gamma$ for different values of capacitors. The noise RMS voltage calculated from \cref{eqn:SC_LP_Act_Vn_total} perfectly fits the simulated noise validating the estimation method.

\section{Conclusion}\label{sec:conclusion}
The design of low-noise SC filters often comes at the cost of higher power consumption. The optimization of SC filters for achieving at the same time low-noise operation at low-power therefore requires an accurate estimation of the integrated noise at the filter output. In Part~I of this paper, we have shown how the original Bode theorem can be extended to active SC circuits using OTAs with capacitive feedback. This generalization allows the calculation of the thermal noise voltage variance across each capacitor of the circuit by simple inspection of several equivalent schematics made of capacitors only, avoiding the evaluation of complex transfer functions and cumbersome integrals. This Part~II presents how the extended Bode theorem can also be applied to SC filters built with OTAs. It is illustrated by three examples, including a passive first-order LP filter (which actually can be calculated with the original Bode theorem), the basic stray-insensitive integrator and finally an active first-order LP filter. The analytical results obtained from the extend Bode theorem are successfully validated using transient noise simulations. The simulations results are very close to the analytical expressions demonstrating the effectiveness of the proposed method also for SC filters based on OTAs.

\section{Appendix}

\subsection{Integrator - Derivation of \Varp{V}{\aC}{2}}\label{sec:SC_int_V2naCp2}
The variance of the noise voltage across $\aC$ at the end of \phase{2} can be evaluated using the extended Bode theorem \cref{eqn:V2n_Bode_OTA} with the help of the schematics of \cref{fig:SC_int_ph2abcd} for the evaluation of \Cinf{}, \Cinfp{} and \Co{} and of the feedback gain given by
\begin{equation}\label{eqn:SC_int_hfb}
  \hfb = \frac{V}{V_{out}} = \frac{1}{1+\a+\ain},
\end{equation}
where $\ain \triangleq C_{in}/C$. This leads to
\begin{equation}\label{eqn:SC_int_V2naCfull}
  \Varp{V}{\aC}{2} = \frac{\kT}{\aC} \cdot \frac{\aL (1+\ain)+\ain+\gamma \cdot \a}{\aL(1+\ain)+\a+\ain},
\end{equation}
where $\aL \triangleq C_L/C$. Usually $\aC$ and $\Cin$ can be considered much smaller than $C$ (i.e. $\a \ll 1$ and $\ain \ll 1$) leading to
\begin{equation}\label{eqn:SC_int_V2naCa1}
  \Varp{V}{\aC}{2} \cong \frac{\kT}{\aC} \cdot \frac{\aL+\ain+\gamma \cdot \a}{\aL+\a+\ain}.
\end{equation}

\subsection{Integrator - Derivation of direct noise}\label{sec:SC_int_direct}
The output noise voltage variance at the end of \phase{1} due to the direct noise can be calculated using the extended Bode theorem \cref{eqn:V2n_Bode_OTA} applied to capacitor $\CL$. Using of the schematics of \cref{fig:SC_int_ph1abcd} this results in
\begin{equation}\label{eqn:SC_int_V2nCLfull}
  \Varp{V}{\CL}{1} = \frac{\gamma \cdot \kT}{C} \cdot \frac{(1+\a+\ain)(1+\ain)}{\ain+\aL(1+\ain)},
\end{equation}
which for $\a \ll 1$ and $\ain \ll 1$ reduces to
\begin{equation}\label{eqn:SC_int_V2nCLapprox}
  \Varp{V}{\CL}{1} \cong \frac{\gamma \cdot \kT}{\CL + \Cin}.
\end{equation}

\subsection{OTA-based LP active filter - Derivation of \Varp{V}{C}{2}}\label{sec:SC_LP_Act_V2nCp2}
The variance of the noise voltage across $C$ at the end of \phase{2} can be calculated using the extended Bode theorem \cref{eqn:V2n_Bode_OTA}. Capacitances \Cinf{}, \Cinfp{} and \Co{} are calculated with the help of \cref{fig:SC_LP_Act_ph2abcd}. \Cref{eqn:V2n_Bode_OTA} also requires the feedback gain $\hfb$ during \phase{2} which for $C_1=C_2=\aC$ is given by
\begin{equation}\label{eqn:SC_LP_Act_hfb}
  \hfb = \frac{V}{V_{out}} = \frac{C+C_2}{C + C_1 + C_2 + C_{in}} = \frac{1+\alpha}{1+2 \alpha + \ain}.
\end{equation}
Applying the extended Bode theorem \cref{eqn:V2n_Bode_OTA} and separating the contribution of the OTA and the switches leads to
\begin{equation}\label{eqn:SC_LP_Act_2nCp2}
  \Varp{V}{C}{2} = \frac{\kT}{C} \cdot \left(\gamma \cdot \botap{2} + \bswip{2}\right),
\end{equation}
where \botap{2} and \bswip{2} are given by \cref{eqn:SC_LP_Act_beta_ota} and \cref{eqn:SC_LP_Act_beta_sw2}, respectively.
\setcounter{equation}{55}
For $\a \ll 1$ and $\ain \ll 1$, \cref{eqn:SC_LP_Act_beta_ota} and \cref{eqn:SC_LP_Act_beta_sw2} reduce to
\begin{subequations}
  \begin{align}
    \botap{2} &\cong \frac{(\a+\ain)^2}{\aL+\a+\ain},\label{eqn:SC_LP_Act_beta_ota_approx}\\
    \bswip{2} &\cong \a \cdot \left(1+\frac{\aL^2}{(\aL+\ain)(\aL+\a+\ain)}\right),\label{eqn:SC_LP_Act_beta_sw_approx}
  \end{align}
\end{subequations}

\bibliographystyle{IEEEtran}


\bibliography{References}


\begin{IEEEbiography}[{\includegraphics[width=1in,height=1.25in,clip,keepaspectratio]{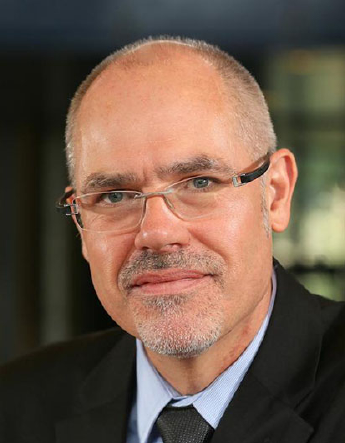}}]{Christian Enz}
(M’84, S'12) received the M.S. and Ph.D. degrees in Electrical Engineering from the EPFL in 1984 and 1989 respectively. He is currently Professor at EPFL, Director of the Institute of Microengineering and head of the IC Lab. Until April 2013 he was VP at the Swiss Center for Electronics and Microtechnology (CSEM) in Neuch\^{a}tel, Switzerland where he was heading the Integrated and Wireless Systems Division. Prior to joining CSEM, he was Principal Senior Engineer at Conexant (formerly Rockwell Semiconductor Systems), Newport Beach, CA, where he was responsible for the modeling and characterization of MOS transistors for RF applications. His technical interests and expertise are in the field of ultralow-power analog and RF IC design, wireless sensor networks and semiconductor device modeling. Together with E. Vittoz and F. Krummenacher he is the developer of the EKV MOS transistor model. He is the author and co-author of more than 250 scientific papers and has contributed to numerous conference presentations and advanced engineering courses. He is an individual member of the Swiss Academy of Engineering Sciences (SATW). He has been an elected member of the IEEE Solid-State Circuits Society (SSCS) AdCom from 2012 to 2014. He is also the Chair of the IEEE SSCS Chapter of Switzerland.
\end{IEEEbiography}


\begin{IEEEbiography}[{\includegraphics[width=1in,height=1.25in,clip,keepaspectratio]{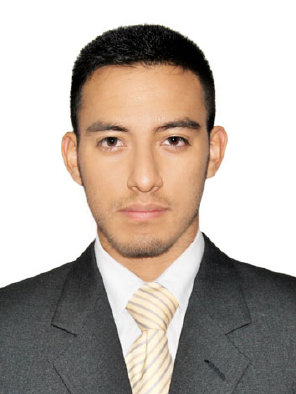}}]{Sammy Cerida Rengifo}
received the B.S. degree, with honors, in electrical engineering from Pontifical Catholic University of Peru (PUCP) in 2014. He obtained the M.S. joint degree in Micro and Nanotechnologies for Integrated Systems (Master Nanotech) between Politecnico di Torino, Grenoble INP, and EPFL. His master thesis research was carried out at EM Microelectronic-Marin, Neuch\^{a}tel, Switzerland, on the investigation and analysis of next-generation UHF RFID tags. In 2017, he joined the CSEM in Neuch\^{a}tel, where he is currently working towards his PhD in the field of ultra low-power radars.
\end{IEEEbiography}


\begin{IEEEbiography}[{\includegraphics[width=1in,height=1.25in,clip,keepaspectratio]{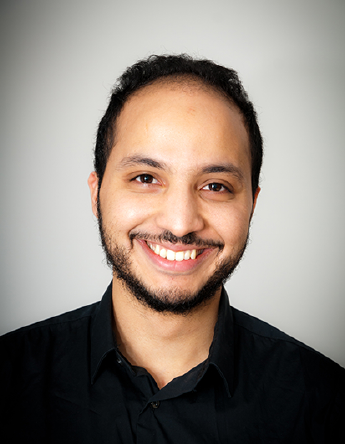}}]{Assim Boukhayma}
received the graduate engineering degree (D.I.) in information and communication technology and the M.Sc. in microelectronics and embedded systems architecture from Institut Mines Telecom (IMT Atlantique), France, in 2013. He was awarded with the graduate research fellowship for doctoral studies from the French atomic energy commission (CEA) and the French ministry of defense (DGA). He received the Ph.D. from EPFL in 2016 on the topic of Ultra Low Noise CMOS Image Sensors. In 2017, he was awarded the Springer Theses prize in recognition of outstanding Ph.D. research. He is currently a scientist at EPFL ICLAB, conducting research in the areas of image sensors and noise in circuits and systems.
From 2012 to 2015, he worked as a researcher at Commissariat a l’Energie Atomique (CEA-LETI), Grenoble, France. From 2011 to 2012, he worked with Bouygues-Telecom as a Telecommunication Radio Junior Engineer.
\end{IEEEbiography}


\begin{IEEEbiography}[{\includegraphics[width=1in,height=1.25in,clip,keepaspectratio]{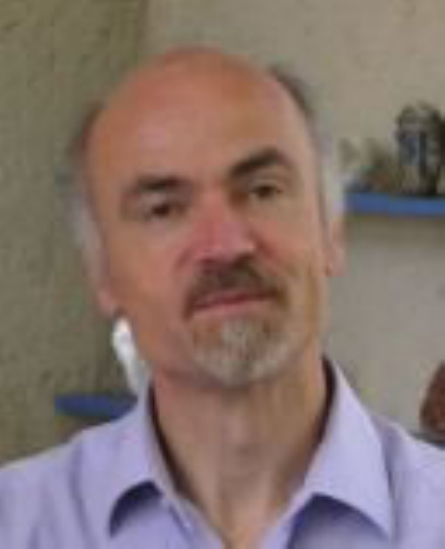}}]{Fran\c{c}ois Krummenacher}
received the M.S. and Ph.D. degrees in electrical engineering from the Swiss Federal Institute of Technology (EPFL) in 1979 and 1985 respectively. He has been with the Electronics Laboratory of EPFL since 1979, working in the field of low-power analog and mixed analog/digital CMOS IC design, as well as in deep sub-micron and high-voltage MOSFET device compact modeling. Dr. Krummenacher is the author or co-author of more than 120 scientific publications in these fields. Since 1989 he has also been working as an independent consultant, providing scientific and technical expertise in IC design to numerous local and international industries and research labs.
\end{IEEEbiography}

\end{document}